\documentclass[pre,twocolumn,amsmath,amssymb,floats,superscriptaddress,nofootinbib,10pt]{revtex4-2}

\pdfoutput=1

\AtBeginDocument{%
    \newwrite\bibnotes
    \def\bibnotesext{Notes.bib}
    \immediate\openout\bibnotes=\jobname\bibnotesext
    \immediate\write\bibnotes{@CONTROL{REVTEX42Control}}
    \immediate\write\bibnotes{@CONTROL{%
    apsrev41Control,author="08",editor="1",pages="1",title="0",year="1"}}
     \if@filesw
     \immediate\write\@auxout{\string\citation{apsrev42Control}}%
    \fi
}%

\usepackage{graphicx}
\usepackage{dcolumn}
\usepackage{bm}
\usepackage{revsymb}
\usepackage[usenames]{color}
\usepackage{subfigure}
\usepackage{color}
\usepackage{physics}
\usepackage{amsmath}
\usepackage[usenames]{color}
\usepackage{amsfonts}
\usepackage{graphicx}
\usepackage{dcolumn}
\usepackage{bm}
\usepackage{revsymb}
\usepackage[usenames]{color}
\usepackage{amssymb,bbm}

\usepackage{xcolor}

\newcommand{\bea}{\begin{eqnarray}}
\newcommand{\eea}{\end{eqnarray}}
\newcommand{\rmd} {{\rm d}}

\newcommand{\rmi} {{\rm i}}

\definecolor{nblue}{RGB}{28,130,185}

\definecolor{cgreen}{RGB}{76,153,0}

\definecolor{myorange}{RGB}{245,156,74}

\definecolor{ogreen} {RGB}{71,191,145}

\usepackage{hyperref}
\hypersetup{
  colorlinks=true,
  citecolor=magenta,
  urlcolor=-myorange
}

\usepackage{hyperref}
\hypersetup{
  colorlinks=true,
  citecolor=magenta,
  urlcolor=-myorange
}

\newcommand{\quotes}[1]{``#1''}

\usepackage{quoting}

\makeatletter
\newcommand\thankssymb[1]{\textsuperscript{a}}
\makeatother

\newcommand{\myparagraph}[1]{\medskip\noindent\textit{#1.---}}

\begin{document}

\title{Lift force in odd compressible fluids}

\author{Ruben Lier
\thankssymb{1}}
\email{rubenl@pks.mpg.de}
\affiliation{Max Planck Institute for the Physics of Complex Systems, 01187 Dresden, Germany}
\affiliation{Würzburg-Dresden Cluster of Excellence ct.qmat,  01187 Dresden, Germany}

\author{Charlie Duclut
\thankssymb{1}}
\email{charlie.duclut@curie.fr}
\affiliation{Laboratoire Physico-Chimie Curie, UMR 168, Institut Curie, PSL Research University, CNRS, Sorbonne Université, 75005 Paris, France}
\affiliation{Universit\'e Paris Cit\'e, Laboratoire Mati\`ere et Syst\`emes Complexes (MSC), UMR 7057 CNRS,  F-75205 Paris,  France}
\affiliation{Max Planck Institute for the Physics of Complex Systems, 01187 Dresden, Germany}

\author{Stefano Bo}
\email{stefano.bo@kcl.ac.uk}
\affiliation{Max Planck Institute for the Physics of Complex Systems, 01187 Dresden, Germany}
\affiliation{Department of Physics, King's College London, London WC2R 2LS, U.K.}

\author{Jay~Armas}
\email{j.armas@uva.nl}
\affiliation{Institute for Theoretical Physics, University of Amsterdam, 1090 GL Amsterdam, The Netherlands}
\affiliation{Dutch Institute for Emergent Phenomena (DIEP), University of Amsterdam, 1090 GL Amsterdam, The Netherlands}

\author{Frank J\"{u}licher}
\email{julicher@pks.mpg.de}
\affiliation{Max Planck Institute for the Physics of Complex Systems, 01187 Dresden, Germany}
\affiliation{Center for Systems Biology Dresden, Pfotenhauerstrasse 108, 01307 Dresden, Germany}
\affiliation{Cluster of Excellence Physics of Life, TU Dresden, 01062 Dresden, Germany}
\author{Piotr Sur\'{o}wka}
\email{piotr.surowka@pwr.edu.pl}
\affiliation{Institute of Theoretical Physics, Wrocław University of Science and Technology, 50-370 Wrocław, Poland}
\affiliation{Institute for Theoretical Physics, University of Amsterdam, 1090 GL Amsterdam, The Netherlands}
\affiliation{Dutch Institute for Emergent Phenomena (DIEP), University of Amsterdam, 1090 GL Amsterdam, The Netherlands}

\renewcommand{\thefootnote}{\alph{footnote}}
\footnotetext[1]{These authors contributed equally.}
\renewcommand{\thefootnote}{\arabic{footnote}}

\begin{abstract}
When a body moves through a fluid, it can experience a force orthogonal to its movement, called lift force.
Odd viscous fluids break parity and time-reversal symmetry,
suggesting the existence of an odd lift force on  tracer particles,  even at vanishing Reynolds numbers and for symmetric geometries.  It was previously found that an incompressible odd fluid cannot induce lift force on a tracer particle with no-slip boundary conditions, making signatures of odd viscosity in the two-dimensional bulk elusive.
By computing the response matrix for a tracer particle, we show that an odd compressible fluid can produce an odd lift force. Using \emph{shell localization}, we provide analytic expressions for the drag and odd lift forces acting on the tracer particle in a steady state and also at finite frequency. Importantly, we find that the existence of an odd lift force in a steady state requires taking into account the non-conservation of the fluid mass density due to the coupling between the two-dimensional surface and the three-dimensional bulk fluid.
\end{abstract}

\maketitle

\myparagraph{Introduction.} Odd materials are characterized by the breaking of parity symmetry, which manifest itself in  viscous and elastic tensor contributions that are odd under index exchange. Breaking this symmetry results in the emergence of novel phenomena, endowing odd materials with fascinating properties that are interesting for various fields of physics, including electron fluids \cite{pellegrino2017nonlocal,narozhny2019magnetohydrodynamics,berdyugin2019measuring}, topological waves \cite{fossati2022odd,green2020topological,souslov2019topological,tauber2019bulkinterface}, fluid dynamics~\cite{banerjee2017odd,abanov2018odd,ganeshan2017odd,khain2022stokes}, complex materials~\cite{scheibner2020odd,braverman2021topological,floyd2022signatures,fruchart2022odda,liao2020rectification}, soft active matter, statistical physics and biological physics~\cite{epstein2020timereversal,hargus2020time,lier2022passive,han2021fluctuating,hargus2021odd,soni2019odd,reichhardt2019active,markovich2021odd,reichhardt2022active,furthauer2012active,tan2022odd,banerjee2021active}. 
Notably, these materials are now within experimental reach and their properties can be measured, validated and further explored~\cite{berdyugin2019measuring,soni2019odd,tan2022odd}.

The simplest examples of odd materials are odd fluids, which are characterized by odd viscosity. Odd viscosity is a transport coefficient in two dimensions breaking parity and time-reversal symmetry, which can occur in passive fluids subject to a background magnetic field ~\cite{avron1995viscosity,levay1995berry}, as well as in active chiral systems~\cite{furthauer2012active,banerjee2017odd}.

The signatures of odd viscosity in fluids have been explored in various contexts (see e.g. \cite{avron1998odd,lucas2014phenomenology,kogan2016lift,alekseev2016negative,banerjee2017odd,ganeshan2017odd,scaffidi2017hydrodynamic,delacretaz2017transport,abanov2018odd,gusev2018viscous,soni2019odd,markovich2021odd,hosaka2021nonreciprocal,hargus2021odd,hosaka2021hydrodynamic,khain2022stokes,fruchart2022odd,banerjee2022hydrodynamic,han2021fluctuating}).
The experimental realization of an active odd fluid in Ref.~\cite{soni2019odd} showed  that the strongest signatures of odd behavior, such as edge flow or the rotation of asymmetric droplets, are found at interfaces.
Inserting a tracer or probe particle in an odd fluid naturally introduces a boundary, making it an ideal candidate to probe the odd properties of a fluid, and has been the subject of several numerical and theoretical studies~\cite{reichhardt2019active,reichhardt2022active,kogan2016lift,ganeshan2017odd,hosaka2021nonreciprocal,hosaka2021hydrodynamic}. 
In particular, due to the parity-breaking nature of odd viscosity, symmetry allows a fluid with a constant velocity at infinity not only to induce a drag force on a tracer particle, but also a lift force, orthogonal to the movement of the tracer.
This odd lift force is allowed at vanishing Reynolds number and in a symmetric geometry. This illustrates its different physical origin compared to the lift force observed for instance in aeronautics, that requires a nonvanishing Reynolds number or a symmetry-breaking mechanism such as the shape of the wing \cite{bureau2022lift}.

Surprisingly, such a lift force is absent\footnote{A nonvanishing odd lift force, on a tracer in an incompressible fluid, assuming no-slip boundary conditions, was obtained in Ref.~\cite{kogan2016lift}, but was contradicted in Ref.~\cite{ganeshan2017odd}. The discrepancy can be traced to the computation of the force on the probe, which in Ref.~\cite{kogan2016lift} used an incorrect pressure field.} in \textit{incompressible} odd fluids~\cite{ganeshan2017odd}, and the motion of a tracer particle cannot be used to detect signatures of odd properties in these systems.

This brings us to a variant of the more-than-a-century-old question:
\emph{how much force does a tracer particle in a fluid experience?} Answering this question typically requires finding a smooth and regular solution for the velocity profile of the fluid flows satisfying appropriate boundary conditions near the tracer particle and far away from it. However, when one tries this for two-dimensional fluids one encounters a problem, commonly known as the Stokes paradox, which prevents a solution to the Stokes equation for a disk moving through a two-dimensional fluid with infinitely large system size \cite{lamb1932hydrodynamics}. The Stokes paradox can be circumvented by adding a scale to the problem which \quotes{regularizes} the paradox. One way to do this is, is a through the Oseen approximation \cite{oseen1910uber}, which introduces the far field velocity through an inertia term. The Oseen approximation can be improved in an iterative way, which is called asymptotic expansion \cite{proudman1957expansions,chester_breach_proudman_1969,veysey2007simple,rosenhead1963laminar,happel1983low,vandyke1975perturbation}. Another way that the Stokes paradox is evaded, is by assuming that the two-dimensional fluid is in contact with a three-dimensional bulk, to which momentum is relaxed \cite{saffman1975brownian,saffman1976brownian,seki1993brownian}. This is what will be considered in this work.

In this Letter, we show that a tracer particle in an odd \textit{compressible} fluid experiences a lift force proportional to the odd viscosity coefficient, and that compressibility is a necessary condition for the existence of an odd lift force in two dimensions. As commonly done when studying the motion of tracers in fluids and to make direct contact with the incompressible case studied in Ref.~\cite{ganeshan2017odd}, we consider no-slip boundary conditions on the surface of the tracer.
Lifting the incompressibility constraint dramatically complicates the two-dimensional fluid equations, the description of the fluid velocity requires, in addition to the stream function, a second scalar field. The differential equations of these two scalar fields are coupled due through odd viscosity. In addition, one also needs to account for the non-trivial role of density.
To tackle these difficulties, we avoid computing the fluid profile and instead use the \quotes{shell localization} approach~\cite{levine2001response,camley2011creeping,levine2002dynamics} to analytically compute the drag and lift forces on a tracer particle in two different situations: a fluid in a steady-state configuration, and a fluid excited by an external force with finite driving frequency.

Crucially ---and this point was overlooked in previous studies on this subject in which an instantaneous density relaxation was considered~\cite{hosaka2021nonreciprocal}---
we show that lift force only persists in a steady state in systems for which the density is not conserved.
Non-conservation of density is generic in active systems as a consequence of birth and death processes, for instance in~\quotes{Malthusian flocks}~\cite{toner2012birth,chen2020moving}, cellular tissues~\cite{ranft2010fluidization}, and in chemotactic systems~\cite{benalizinati2022stochastic}.
Furthermore, absence of mass density conservation in two dimensions can arise from exchanges with a three-dimensional fluid bulk~\cite{barentin2000shear,elfring2016surface}.  
This is for instance the case if the odd properties of the fluid stem from the activity of chiral particles, such as bacteria~\cite{petroff2015fastmoving} or spermatozoa~\cite{riedel2005selforganized} that swim in a  three-dimensional fluid and can accumulate at a surface.

As a further step, we also investigate the response of a probe excited periodically. At finite frequency, we show that an odd lift force can be measured in compressible fluids even if the mass density is conserved. This paves the way towards measurements of odd transport coefficients using frequency-dependent micro-rheology.

\myparagraph{Compressible odd fluid}
We consider a thin layer of an odd compressible viscous fluid at the interface between two bulk (even) fluids, for instance water and air. For simplicity, we consider this layer to be flat and infinitely thin, such that the odd fluid can be described effectively as two-dimensional. The stress tensor associated with the mechanical properties of the odd fluid with velocity field $v_i$ reads
\begin{align}
\label{eq_stressTensor}
    \sigma_{ij} = 2 \eta_{\rm s} \partial_{\langle i} v_{j\rangle} + 2\eta_{\rm o} \partial_{\{ i} v_{j\}} + \left( \eta_{\rm b} \partial_k v_k -P \right )\delta_{ij} \, ,
\end{align}
where $i,j$ denote two-dimensional Cartesian coordinates and where summation over repeated indices is implied. For an arbitrary tensor $A_{ij}$, we have introduced the notation $A_{\langle ij \rangle}=(A_{ij} + A_{ji})/2-A_{kk}\delta_{ij}/2$ for its traceless symmetric part, such that $\partial_{\langle i} v_{j\rangle}$ is the fluid shear rate. We have also introduced the odd tensor contraction $A_{\{ ij \}}= (\varepsilon_{ik} A_{kj} + \varepsilon_{ik} A_{jk} + \varepsilon_{jk} A_{ki} + \varepsilon_{jk} A_{ik} )/4$, where $\varepsilon_{ij}$ denotes the fully antisymmetric tensor in two dimensions with $\varepsilon_{12}=-\varepsilon_{21}=1$. Finally, we denote by $\eta_{\rm s,b,o}$ the shear, bulk and odd viscosities of the fluid, and by $P$ its pressure field.

The divergence of the stress tensor~\eqref{eq_stressTensor} then allows us to write the momentum balance equation, which corresponds to the odd version of the Navier--Stokes equation. It reads:
\begin{align}%
\label{eq_oddNavierStokes}
\begin{split}
    \partial_t \pi_i &+  v_k \partial_k \pi_i= \eta_{\rm s} \partial_k\partial_k v_i + \eta_{\rm b} \partial_i \partial_k v_k - \partial_i P \\ 
    & \quad + \eta_{\rm o} \varepsilon_{ij} \partial_k\partial_k v_j
    - \frac{1}{\tau } \pi_i + f_i \, .
\end{split}
\end{align}%
where $\pi_i = \rho v_i $ is the fluid momentum density with $\rho$ the local mass density. 
The first line of Eq.~\eqref{eq_oddNavierStokes} is the usual isotropic Navier--Stokes equation, while the first term in the second line is the signature of odd two-dimensional fluids.
In addition, we have included in Eq.~\eqref{eq_oddNavierStokes} a momentum relaxation process with timescale~$\tau$. This process accounts for linear friction between the two-dimensional fluid and the three-dimensional bulk which can generically exists in our geometry.
The last term $f_i$ in Eq.~\eqref{eq_oddNavierStokes} is an external force density acting on the fluid, which will prove convenient to compute the drag and lift coefficients of a probe immersed in the fluid. 

As will become clear below, the compressibility of the odd fluid layer is a necessary condition to observe a nonvanishing lift force. A compressible fluid can be described by providing an equation of state for the pressure field~$P$ written as a series expansion in powers of the fluid density~$\rho$. For a weakly compressible fluid that we consider here, we keep only the first nontrivial order and write:
\begin{align}
\label{eq_pressure}
P(\rho) = P_0 + \chi \frac{(\rho-\rho_0)}{\rho_0} \, ,    
\end{align}
where $\chi^{-1}$ is the compressibility, and $P_0$ and $\rho_0$ are the reference pressure and density, respectively. Finally, the mass density obeys the balance equation
\begin{align}
    \partial_t \rho + \partial_k (\rho v_k) &= - \frac{1}{\kappa} (\rho-\rho_0) \, , \label{eq_massContinuityEquation}
\end{align}
where we have included a mass exchange process with timescale $\kappa$ to account for particle exchange with the bulk of the fluid \cite{elfring2016surface}. Note that linear terms proportional to the density in Eq.~\eqref{eq_pressure} and  in Eq.~\eqref{eq_massContinuityEquation} would also be allowed in an active fluid layer~\cite{julicher2018hydrodynamic}, such as a cell epithelium. In this specific case, $\chi (\rho-\rho_0)/\rho$ would correspond to an active isotropic stress and $(\rho-\rho_0)/\kappa$ would account for cell divisions and extrusions. Finally, we emphasize that the case of a momentum-conserving, mass-conserving, or incompressible fluid can be easily recovered by taking respectively the limit~$\tau\to\infty$, $\kappa\to\infty$, or $\chi\to\infty$ in Eqs.~\eqref{eq_oddNavierStokes}-\eqref{eq_massContinuityEquation}. These coupled equations thus provides the ideal starting point for studying odd effects in two-dimensional fluid layers.    

To simplify the system of coupled nonlinear differential equations~\eqref{eq_oddNavierStokes}-\eqref{eq_massContinuityEquation}, we linearize it to first order in $v_i$ and $\delta\rho=\rho-\rho_0$ near a vanishing velocity and homogeneous reference state. The balance equations then take the form
\begin{subequations}\label{eqs_compressibleStokes}
\begin{align}%
\begin{split}
    \rho_0\partial_t v_i &= \eta_{\rm s} \partial_k\partial_k v_i + \eta_{\rm b} \partial_i \partial_k v_k \\ 
    & + \eta_{\rm o} \varepsilon_{ij} \partial_k\partial_k v_j
    - \partial_i P  - \frac{\rho_0}{\tau }v_i + f_i \, , 
\end{split}\label{eq_oddStokes}\\
    \partial_t \delta\rho + \rho_0\partial_k  v_k &= - \frac{1}{\kappa} \delta\rho \, . \label{eq_linearMassContinuityEquation}
\end{align}%
\end{subequations}%
We will use these equations to compute the response of a probe to an external force in an odd fluid.

\myparagraph{Shell localization}
Having defined the equations of motion, we move to Fourier space with the convention
\begin{align}
 g (t , x_i) = \frac{1}{(2\pi)^3} \int {\rm d} \omega {\rm d}^2 k \, g(\omega,k_i ) e^{- \rmi \omega t  +  \rmi k_j x_j }     ~~, 
\end{align}
for some function $ g (t , x_i)$ so that Eq.~\eqref{eqs_compressibleStokes} can be written in matrix form $\mathcal{G}_{ij} v_j = f_i$ with $\mathcal{G}_{ij}$ given by
 \begin{align} 
 \begin{split}
    \mathcal{G}_{ij}   = \hat{k}_i  \hat{k}_{j} \left[\frac{\rho_0}{\tau}   - \rmi \omega      \rho_0   + \left( \eta_{\rm s} +  \eta_{\rm b}   +  \frac{ \chi \kappa }{1 - \rmi \omega \kappa }  \right) k^2          \right]   \\ 
        +     (\delta_{ij}- \hat{k}_i  \hat{k}_{j})  \left[\frac{\rho_0}{\tau}  -   \rmi \omega      \rho_0    + \eta_{\rm s} k^2     
     \right]   + 
     \varepsilon_{ij}  \eta_{\rm o }  k^2    ~~   , 
 \end{split}
 \label{eq_fullResponseMatrix}
\end{align}
where $k=\sqrt{k_ik_i}$ and $\hat{k}_i = k_i/k$. This relation can be inverted as
\begin{align}
   v_i (k_i , \omega ) =    \mathcal{M}_{ij } (k_i , \omega )  f_j  (k_i , \omega )      ~~  ,  \label{wihwiuh1iu111}
\end{align}
where we have defined $\mathcal{M}_{ij } = \mathcal{G}^{-1}_{ij} $. Equaqtion~\eqref{wihwiuh1iu111} yields the velocity induced by a force distribution. Specifically, we consider the force applied on a tracer particle, which is a rigid disk of radius $a$ located at the origin. Due to the rotational symmetry of the disk, we can decompose the force density as $f_j (k_i,\omega) =  L(k)  \mathcal{F}_j (\omega)$. The shell localization method consists in considering that the force density is located in real space according to~\cite{levine2001response,camley2011creeping,mackintosh1991orientational}: 
\begin{align}
    L (x) =  \frac{1}{2 \pi a }  \delta ( |x|- a )
    \label{eq_forceDensity_real} ~~ .
\end{align}
Equation~\eqref{eq_forceDensity_real} enforces the force density exerted by the disk on the fluid to be uniformly distributed along the entire edge of the disk. The disk is coupled to the fluid through a no-slip boundary condition, which equates the velocity of the tracer particle to the fluid velocity at the edge of the tracer particle. Fourier transforming Eq.~\eqref{eq_forceDensity_real} yields $L(k) =  J_0  (a k)$ with $J_n(z)$ the $n^{\rm th}$ Bessel function of the first kind. 
The velocity of the disk located at $|x| =0 $  is then directly given by the inverse Fourier transform at the origin
\begin{align}
  v_{i} (|x|=0, \omega)  =   \mathbb{M}_{ij } (\omega )   \mathcal{F}_j (\omega) ~~ , 
\end{align}
where the ``response matrix'' is 
\begin{align}  
\mathbb{M}_{ij } (\omega )  =  \frac{1}{(2 \pi)^2 } \int_0^{2 \pi } \!\! {\rm d} \theta  \int_0^{\infty} \!\! {\rm d} k \, k L (k)  \mathcal{M}_{ij }  (k_i, \omega )  \,  . \label{eq_responseMatrix}
\end{align}
The response matrix $\mathbb{M}_{ij } (\omega )$ encodes the velocity of a rigid probe immersed in an odd fluid as a function of the applied  (frequency-dependent) force $\mathcal{F}_j (\omega)$. Using the disk radius $a$ we can introduce the dimensionless coefficients
\begin{align}
\begin{split}
    z_i   & =  a k_i   \,   ,   \, \,  
    \tilde \omega =  \omega a^2 \rho_0 / \eta_{\rm s}   \,    ,   \, \,  
    \tilde \eta_{\rm o }   =  \eta_{\rm o } / \eta_{\rm s }   \,    ,   \, \,  
    \tilde \eta_{\rm b}    =  \eta_{\rm b} / \eta_{\rm s}  \\     \tilde \tau   &  =  \tau \eta_{\rm s} / (\rho_0  a^2) \,   ,   \, \,  
    \tilde \chi     =    \chi \rho_0 a^2 /\eta_{\rm s}^2     \,  , \, \, 
    \tilde \kappa  =  \kappa \eta_{\rm s}  / (\rho_0 a^2)   \, , 
\end{split}
\end{align}
so that Eq.~\eqref{eq_fullResponseMatrix} turns into
 \begin{align} 
 \begin{split}
       \mathcal{G}_{ij}   = \frac{\eta_{\rm s}}{a^2 } \Big\{ \hat{z}_i  \hat{z}_{j} \left[ \frac{1}{\tilde \tau} - \rmi \tilde  \omega  + \! \left(1 +  \tilde \eta_{\rm b}   +  \frac{\tilde  \chi \tilde  \kappa }{1 -\rmi \tilde  \omega  \tilde \kappa}  \right)\! z^2   \right]   \\ 
  +     (\delta_{ij}- \hat{z}_i  \hat{z}_{j})  \left[\frac{1}{\tilde \tau} -   \rmi \tilde \omega   +   z^2      
     \right]  + 
     \varepsilon_{ij}  \tilde \eta_{\rm o }  z^2 \Big\}     \,   , 
 \end{split}
\end{align}
where $z=\sqrt{z_i z_i}$ and $\hat{z}_i=z_i/z$. Before considering the most general case of a compressible fluid, where a lift force can arise, we first discuss the limiting case of an incompressible odd fluid.
This corresponds to the limit \mbox{$\tilde \chi \rightarrow \infty$}, for which the matrix $\mathcal{M}$ reads
\begin{align}
    \lim_{\chi \rightarrow \infty} \mathcal{M}_{ij }   (z_k ,\tilde  \omega )  =    \frac{a^2 }{\eta_{\rm s}}  \frac{\delta_{ij}- \hat{z}_i  \hat{z}_{j} }{  z^2 + \tilde \tau^{-1}  - \rmi \tilde \omega}   ~~ .     \label{eq_M_incompressibleLimit}
\end{align}
It may be observed that this matrix is transverse to the wave-vector, indicating the absence of an odd lift force as expected for an incompressible odd fluid~\cite{ganeshan2017odd}. In addition, the odd viscosity transport coefficient is absent, indicating that the response of the tracer particle in the case of an odd incompressible fluid is identical to the response in the case of an even incompressible fluid. In App.~\ref{sec:incompressible_drag}, we verify that in the incompressible case the shell localization gives a drag force that is consistent with results found by explicitly solving the boundary value problem in two instances. Specifically, we recover the result for two-dimensional oscillatory drag \cite{williams1972oscillating,dolfo2020stokes} as well as the result for the drag force found in the Saffman-Delbrück model \cite{saffman1975brownian,saffman1976brownian}, provided we appropriately match the relaxation time to the coefficients of this model~\cite{seki1993brownian}.

\myparagraph{Odd lift force}
We now address the general case of a compressible fluid. In this setting, the response matrix can written as 
\begin{align}
    \mathbb{M}_{ij } (\omega ) =  \frac{1}{ \eta_{\rm s} } ( M_{\rm d} \delta_{ij} - M_{\rm l} \varepsilon_{ij} )\, ,
\end{align}
where $M_{\rm d}$ and $M_{\rm l}$ are respectively the dimensionless response functions for drag force, and for lift force, specific to compressible odd fluids.

\begin{figure}
    \centering
    \includegraphics[width=0.9\columnwidth]{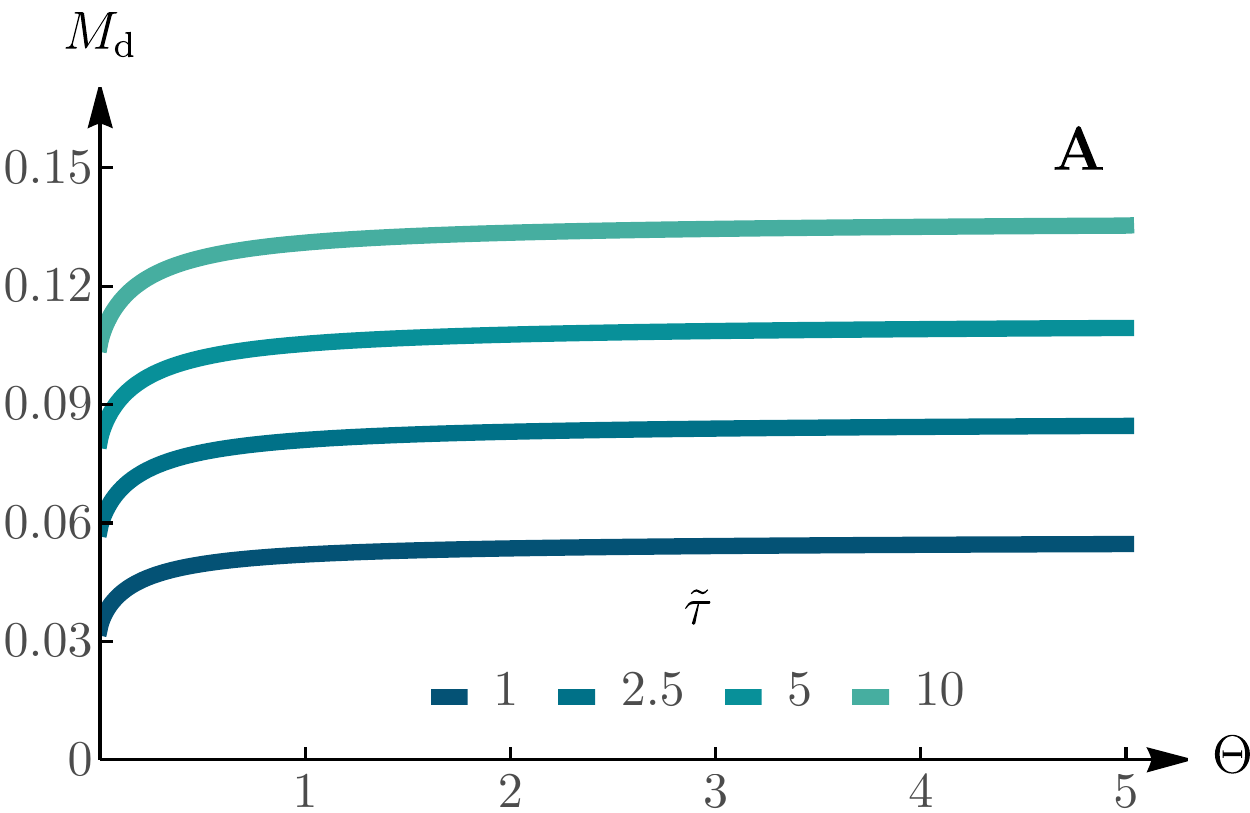}
    \includegraphics[width=0.9\columnwidth]{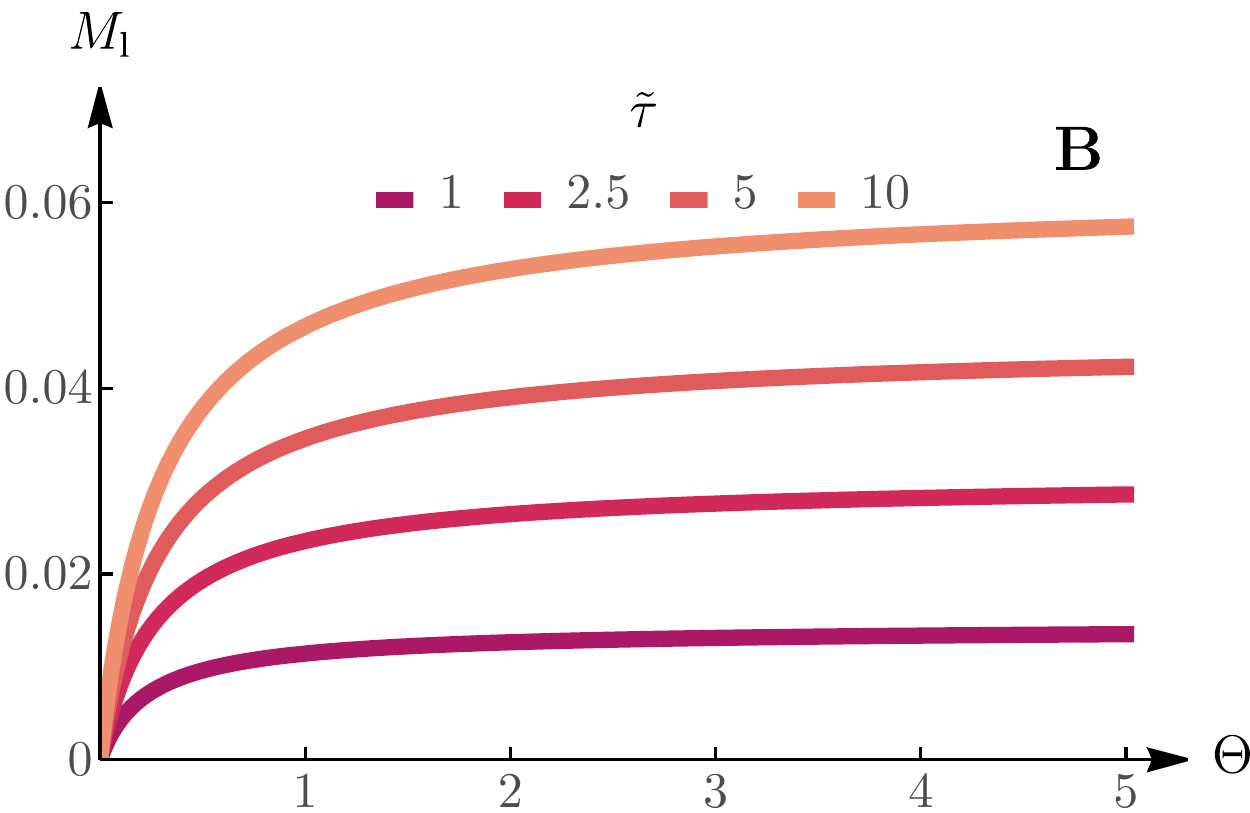}
    \caption{Steady-state drag coefficient $M_{\rm d}$ (\textbf{A}) and lift coefficient $M_{\rm l}$ (\textbf{B}) as a function of the dimensionless inverse compressibility $\Theta=(\tilde\chi\tilde\kappa)^{-1}$ for different values of the relaxation time $\tilde \tau$ and for $\tilde\eta_{\rm b}=\tilde\eta_{\rm o}=1$.}
    \label{fig_steady}
\end{figure}

\myparagraph{Steady-state odd lift force}    
We first consider the steady-state case $\tilde \omega \rightarrow 0$ with a non-vanishing relaxation rate $\tilde \tau^{-1} \neq0$. The drag and lift are obtained by computing the momentum integrals
\begin{subequations}%
\begin{align}%
    M_{\rm d} &= \frac{1}{4\pi} \int {\rm d}z\, J_0(z) \frac{ D(z)}{Q(z)} \, , \\ 
    M_{\rm l} &= \frac{1}{2\pi} \int {\rm d}z\, J_0(z) \frac{L(z)}{Q(z)} \, ,
\end{align}\label{eq_dragLift_steadyState}%
\end{subequations}%
where we have defined:
\begin{subequations}%
\begin{align}%
\begin{split}
    Q(z) &= \tilde \tau ^2 z^4 \left( \tilde\eta_{\rm b} +\tilde \eta_{\rm o} ^2+\Theta^{-1}+1\right) \\
    & \,\,\,\,\, +\tilde \tau  z^2 (\tilde \eta_{\rm b} +\Theta^{-1}+2)+1 \, ,
\end{split}\\
    D(z)&=\tilde \tau  z \left(\tilde \tau  z^2 (\tilde \eta_{\rm b} +\Theta^{-1} +2)+2\right) \, , \,\, L(z)=\tilde \eta_{\rm o}  \tilde \tau ^2 z^3 \, ,
\end{align}%
\end{subequations}%
and where $\Theta^{-1}=\tilde\kappa \tilde \chi$.
As advertised in the introduction, we note that the odd lift force vanishes for $\tilde \eta_{\rm o} \rightarrow 0$, which is expected as it can only be induced by a parity-odd coefficient.
Furthermore, $M_{\rm l}$ is only non-vanishing when $\tilde \kappa^{-1}$ is non-vanishing, since in the steady case the limit $\tilde \kappa \rightarrow \infty$ is equivalent to the incompressible limit for which was shown in Eq.~\eqref{eq_M_incompressibleLimit} that the lift force vanishes.
This means that in a steady state there can only be lift forces when density is not conserved, for instance if exchanges between the surface and three-dimensional fluid, parameterized by the relaxation time $\kappa$, take place.

As we detail in App.~\ref{sec:residues}, where we use Ref.~\cite{lin2013infinitea}, the momentum integrals can be computed analytically using residues but their expression can become lengthy. For the purpose of clarity, we consider a series expansion in powers of the odd viscosity $\tilde \eta_{\rm o}$ and keep the first non-vanishing contribution. Specifically, we find
\begin{subequations}%
\begin{align}%
M_{\rm d } &= \frac{K_0(\tilde\tau^{-1/2}) + K_0[(\Xi\tilde\tau)^{-1/2}]/\Xi }{4\pi}
 + \mathcal{O}(\tilde\eta_{\rm o}^2) \, , \\
M_{\rm l}  &= 
\frac{\tilde\eta_{\rm o} \left[ K_0(\tilde\tau^{-1/2}) - K_0[(\Xi\tilde\tau)^{-1/2}]/\Xi \right]}{2\pi (\Xi-1)}
+ \mathcal{O}(\tilde\eta_{\rm o}^2) \, , 
\end{align}\label{eq_analytical_steady}%
\end{subequations}%
with $\Xi = 1+ \tilde\eta_{\rm b}+\Theta^{-1} $ and where $K_n(x)$ is the $n^{\rm th}$ modified Bessel function of the second kind. In the incompressible fluid limit or for a compressible fluid without mass density relaxation ($ \Theta\to 0$), we find $M_{\rm l} =0$ and $M_{\rm d}= K_0(\tilde\tau^{-1/2})/(4\pi)$.

We now evaluate $M_{\rm d}$ and $M_{\rm l}$ from Eq.~\eqref{eq_dragLift_steadyState} as a function of $\Theta$  and provide the result in Fig.~\ref{fig_steady}. We take $\tilde \eta_{\rm b} = \tilde \eta_{\rm o} =1 $ for the dimensionless viscosities.
We observe in Fig.~\ref{fig_steady} \textbf{A} that the drag force is significantly affected by the momentum relaxation time $\tilde \tau $ but only weakly depends on the compressibility parameter $\Theta$. On the other hand, Fig.~\ref{fig_steady} \textbf{B} shows the crucial role of the compressibility in the magnitude of the lift force, which vanishes in the incompressible limit $\Theta \rightarrow 0$. 

We also consider the limit $\Theta \rightarrow \infty$ which corresponds to an infinitely compressible fluid ($\tilde\chi=0$), or to a fluid with an instantaneous density relaxation $\kappa=0$). In this limit, any deviation from the reference density $\rho_0$ is instantly relaxed to the bulk, such that pressure is constant and plays no role in the response matrix. In this case, our equations reduce to the ones considered in Ref.~\cite{hosaka2021nonreciprocal} where numerical expressions for the response function are computed.

Lastly, we note that the odd lift coefficient $M_{\rm 1}$ can become negative for small values of $\tilde \tau $ and large values of $\Theta$. However, this regime in parameter space for which $\tau\ll1$ is precisely the regime in which momentum relaxation dominates and the system given by Eqs.~\eqref{eqs_compressibleStokes} no longer provides an accurate description of two-dimensional fluid flows.

\begin{figure}
    \centering
    \includegraphics[width=\columnwidth]{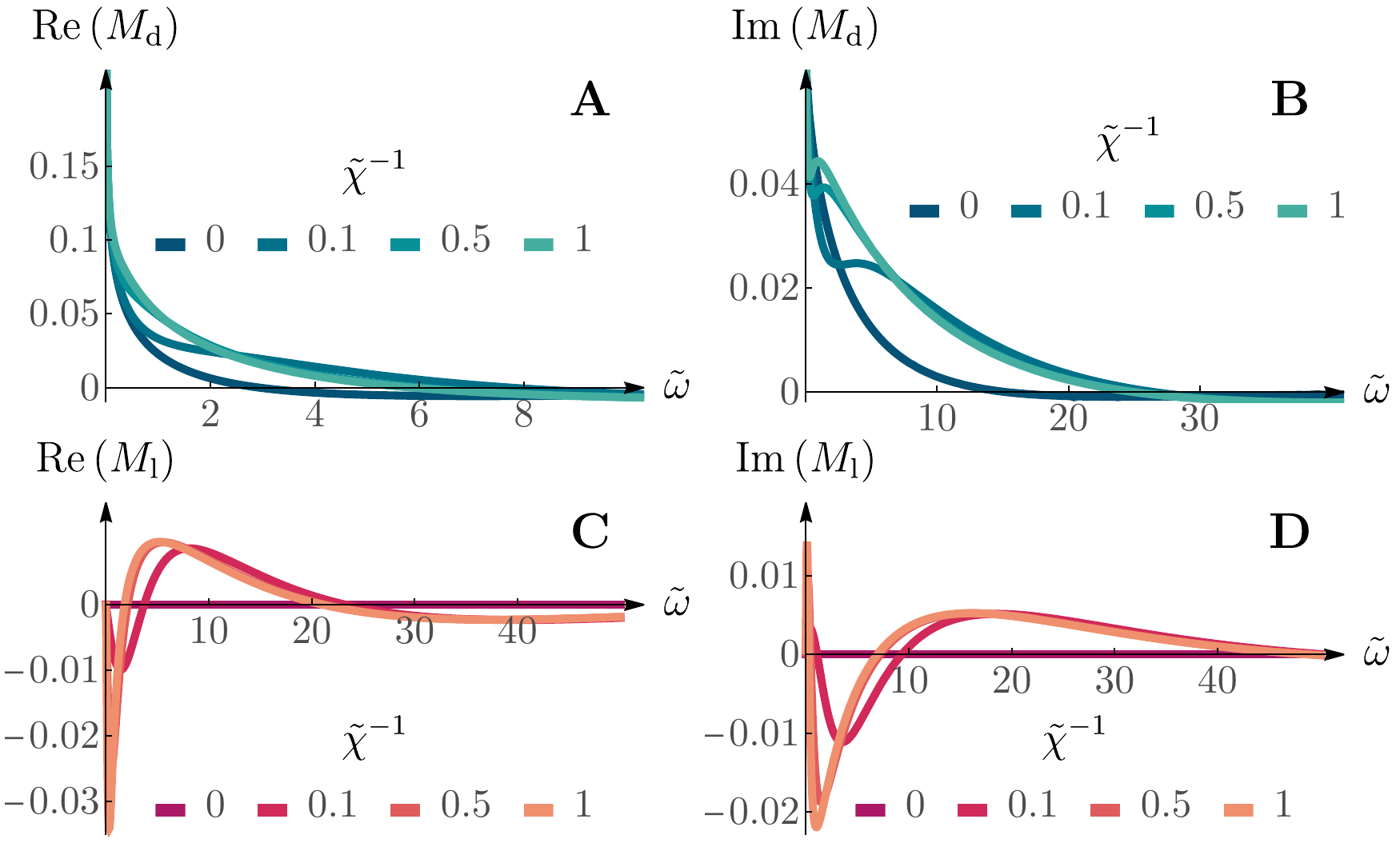}
    \caption{Real (\textbf{A,C}) and imaginary parts (\textbf{B,D}) of the complex drag and lift coefficients $M_{\rm d,l}$ as a function of the dimensionless frequency $\tilde\omega$ for different values of the inverse compressibility $\tilde \chi$ and for $\tilde\eta_{\rm b}=\tilde\eta_{\rm o}=1$.}
    \label{fig_frequency}
\end{figure}

\myparagraph{Frequency-dependent lift force}
We now consider the system in the absence of relaxation processes ($\tilde\tau^{-1}\to0$ and $\tilde\kappa^{-1}\to0$) to focus on the frequency-dependent response of the tracer.
In Fig.~\ref{fig_frequency}, we display the real and imaginary parts of the drag coefficient $M_{\rm d} (\tilde \omega)$ and odd lift coefficient $M_{\rm l} (\tilde \omega)$ as a function of the dimensionless frequency $\tilde\omega$ and for different values of the inverse compressibility $\tilde\chi^{-1}$. The drag coefficient $M_{\rm d} (\tilde \omega)$ diverges as $\tilde\omega\to0$, which is a signature of the Stokes paradox, see Figs.~\ref{fig_frequency}~{\textbf{A}} and {\textbf{B}}. 

On the other hand, the lift coefficient $M_{\rm l} (\tilde \omega)$ vanishes at steady state, see Figs.~\ref{fig_frequency}~{\textbf{C}} and {\textbf{D}}. At finite excitation frequency and compressibility, a nonvanishing odd response can be measured.
Note that both the drag and lift responses vanish at large frequencies, as expected for a fluid.

Additionally, a simple analytic expression for the drag and odd lift coefficient $M_{\rm d,l}$ can be obtained by expanding Eq.~\eqref{eq_responseMatrix} in the absence of relaxation processes ($\tilde\tau^{-1}\to0$ and $\tilde\kappa^{-1}\to0$) and at leading order in the inverse compressibility $\tilde\chi^{-1}$. One obtains
\begin{subequations}%
\begin{align}%
    M_{\rm d} &= \frac{1}{4\pi} K_0 \left(\sqrt{\tilde\omega/\rmi} \right) + \mathcal{O}(\tilde\chi^{-1}) \, , \\
    M_{\rm l} &= \frac{-\rmi \tilde\omega \tilde\eta_{\rm o}}{2\pi\tilde\chi}  K_0 \left(\sqrt{\tilde\omega/\rmi} \right) + \mathcal{O}(\tilde\chi^{-2}) \, .
\end{align}\label{eq_analytic_frequency}%
\end{subequations}%
The drag and lift coefficients have a completely different behavior in the limit of small frequencies. Indeed, we have the expansion\footnote{Note that because the drag coefficient diverges in the limit \mbox{$\tilde\omega\to0$}, the expansion in a series of $\tilde\omega$ must be performed after computing the momentum integral over $z$.}
\begin{subequations}%
\begin{align}%
    M_{\rm d} &= -\frac{1}{8\pi}\left(\!
    \log \frac{\tilde\omega}{4} + 2\gamma_{\rm EM}  - \frac{\rmi \pi}{2}
    \!\right) + \mathcal{O}(\tilde\chi^{-1},\tilde\omega) \, , \\
    M_{\rm l} &= \frac{\rmi \tilde \omega \tilde\eta_{\rm o}}{4\pi\tilde\chi} \left(\!
    \log \frac{\tilde\omega}{4}  + 2\gamma_{\rm EM}  - \frac{\rmi \pi}{2}
    \!\right) + \mathcal{O}(\tilde\chi^{-2},\tilde\omega^2)  \, ,
\end{align}\label{eq_analytic_frequency_expanded}%
\end{subequations}%
which shows a $\log\tilde\omega$ divergence of the drag, as expected from the Stokes paradox, while the odd lift coefficient vanishes as $\tilde\omega\log\tilde\omega$. This difference in the small $\tilde\omega$ behavior is clearly visible in Fig.~\ref{fig_frequency}.

\myparagraph{Discussion}
In this Letter we obtained analytical expressions for the drag and lift coefficients of a disk in a two-dimensional odd compressible fluid. We used a shell localization approach~\cite{levine2001response,camley2011creeping} to study the probe response both at steady-state and at finite frequency. 
In the incompressible limit, we confirmed the absence of odd effects on the tracer with no-slip boundary conditions~\cite{ganeshan2017odd}. 
Having in mind a two-dimensional system embedded in a three-dimensional bulk, we have considered a finite momentum relaxation due to friction, which remedies the Stokes paradox. We found that in order for lift force to be non-vanishing in the steady case, an additional density relaxation due to exchanges with the bulk is required\footnote{Note that in Ref.~\cite{hosaka2021nonreciprocal}, the odd lift force was computed in the limit of a vanishing density relaxation time ($\kappa \rightarrow 0$) using the Lorentz reciprocal theorem~\cite{masoud2019reciprocal}. However, this theorem relies on the index exchange symmetry $\eta_{ij k l }= \eta_{k l ij }$ of the viscosity tensor, which does not hold for an odd fluid. After completing this work, a work appeared where a modified version of the Lorentz reciprocal theorem is introduced which can accommodate for the anti-reciprocal odd viscosity \cite{hosaka2023lorentz} and can therefore be used to overcome this problem.}.

The shell localization approach has also been used to compute drag force for the incompressible Oseen equation \cite{WEISENBORN1984191}. An interesting question is whether it is possible to also apply this computation for the case where the Oseen approximation is applied to odd compressible fluids.
Furthermore, it would be interesting to see whether \quotes{effective boundary conditions}~\cite{korneev2021chiral} accounting for a small finite compressibility can be used to capture the odd lift force on the probe while using an incompressible model in the bulk.  
Finally, when the tracer is excited at finite frequency $\omega$, we found that an odd lift response exists at finite frequency, and vanishes as $\omega \log (\omega )$ in the limit of small frequency. For comparison, the drag response diverges in the same limit as $\log (\omega )$, a signature of Stokes paradox~\cite{williams1972oscillating,dolfo2020stokes}. These results suggest that active micro-rheology could be used to measure the properties of odd viscoelastic materials.

\myparagraph{Acknowledgements}
J.A. is partly supported by the Nederlandse Organizatie voor Wetenschappelijk Onderzoek (NWO) through the NWA Startimpuls funding scheme and by the Dutch Institute for Emergent Phenomena (DIEP) cluster at the University of Amsterdam. R.L. was supported, in part, by the cluster of excellence ct.qmat (EXC 2147, project-id 39085490). P.S. acknowledges the support of the Narodowe Centrum Nauki (NCN) Sonata Bis Grant No. 2019/34/E/ST3/00405 and NWO Klein grant via NWA route 2. 
C.D. acknowledges the support of the LabEx “Who~Am~I?” (ANR-11-LABX-0071) and of the Université Paris Cité IdEx (ANR-18-IDEX-0001) funded by the French Government through its “Investments for the Future” program.
The authors wish to thank Alexandre G. Abanov for insightful comments on the manuscript.

\appendix

\section{Incompressible drag force}
\label{sec:incompressible_drag}

In this Appendix, we explicitly compute the response matrix in two simple incompressible scenarios. In the absence of relaxation ($\tilde\tau^{-1}\to0$), the incompressible response matrix is given by
\begin{align}
\begin{split}
  \mathbb{M}_{ij } (\tilde \omega )   \! &=  \frac{\delta_{ij}}{4 \pi \eta_{\rm s}} \int {\rm d}z \,      \frac{  z  J_0  ( z )      }{  z^2  - \rmi \tilde \omega}  \\ 
  & \!=  -\frac{\delta_{ij}}{8 \pi \eta_{\rm s}}  \!\left[  \log ( \frac{\tilde \omega}{4}   ) +  2 \gamma_{\rm EM} \!  - \rmi \frac{\pi}{2} \!  \right] \!+ \mathcal{O} (\tilde \omega  )      \label{eq_matrix_incompressible_noRelaxation} \, ,    
\end{split}
\end{align}
where $\gamma_{\rm EM}$ is the Euler-Mascheroni constant. We see that this result is divergent in the steady-state limit ($\tilde \omega \rightarrow 0$), which is a signature of the Stokes paradox~\cite{veysey2007simple}. Note that the shell localization result given in Eq.~\eqref{eq_matrix_incompressible_noRelaxation} matches the drag force that one would obtain from solving explicitly the Stokes equation with no-slip boundary conditions~\cite{williams1972oscillating,dolfo2020stokes}. 

A second scenario is the steady-steady case $\tilde \omega \rightarrow 0$  with a finite relaxation rate $\tilde \tau^{-1} \neq0$. The incompressible response matrix now reads
\begin{align}
\begin{split}
\label{eq_matrix_incompressible_steady}
\mathbb{M}_{ij } (0  )   & =  \frac{\delta_{ij}}{4 \pi \eta_{\rm s}} \int {\rm d}z \,    \frac{  z  J_0  ( z )      }{  z^2  + 1/\tilde \tau} \\ 
& =   
\frac{\delta_{ij}}{4 \pi \eta_{\rm s}} \left[ \log (2 \sqrt{\tilde \tau}) -   \gamma_{\rm EM} \right] \! + \mathcal{O} (\tilde \tau^{-1} )  ~ . 
\end{split}
\end{align}
In this case, the response matrix is non-divergent thanks to the momentum relaxation circumventing the Stokes paradox~\cite{evans1988translational,seki1993brownian,ramachandran2010drag,lucas2017stokes}. The result in Eq.~\eqref{eq_matrix_incompressible_steady} can be compared to the result from works of Saffman and Delbrück \cite{saffman1975brownian,saffman1976brownian}, if one matches the relaxation $\tilde \tau$ as~\cite{seki1993brownian}
\begin{align}
    \tilde \tau = \left( \frac{ \eta_{\rm s}}{2 a  \eta^{\prime }_{\rm s}} \right)^2  , \label{eufheiu10909}
\end{align}
where $\eta^{\prime}_{\rm s}$ is the shear viscosity of the surrounding bulk fluid that is tied to the substrate in Refs.~\cite{saffman1975brownian,saffman1976brownian}\footnote{Note in  Refs.~\cite{saffman1975brownian,saffman1976brownian} the shear viscosity in the substrate $ \eta^{(SD)}_{\rm s}$ is three-dimensional and therefore it has different units from the $\eta_{\rm s}$ appearing in this letter.   In Eq.~\eqref{eufheiu10909} the two viscosities are related by  taking $ \eta^{(SD)}_{\rm s} \rightarrow  \eta_{\rm s} / h $, with $h$ being the height of the substrate.}. 
We thus find that in these two instances, the shell localization approach yields the same results as in previous works where the fluid velocity profile is computed over the entire two-dimensional surface \cite{saffman1975brownian,saffman1976brownian}.

\section{Analytical computation of the response matrix}
\label{sec:residues}
In this Appendix, we show how the integrals performed throughout this Letter can be performed using the method of residues.
For a compressible fluid as described in the main text, the response coefficients are obtained by performing momentum integrals that take the form
\begin{align}
    I[R] = \int_0^\infty \rmd z \, R(z) J_0(z) \, ,
\end{align}
where $R(z)=A(z)/B(z)$ is an odd function of $z$, and where $A$ and $B$ are polynomials in $z$. We call $z_n$ the $n^{\rm th}$ root of $B(z)$, such that $B(z_n)=0$. Following Ref.~\cite{lin2013infinite}, the integral $I[R]$ can be computed analytically in terms of the Hankel functions of the first kind~$H^{(1)}_\nu$ and the Bessel functions of the second kind~$Y_\nu$. It reads:
\begin{align}
\begin{split}
    I[R] &= \rmi \pi \sum_{z_n\in \mathbb{C}^+\backslash \mathbb{R}} {\rm Res} \left(R(z) H^{(1)}_0(z),z_n \right) \\
    & \quad - \pi \sum_{z_n\in \mathbb{R}^+} {\rm Res} \left(R(z) Y_0(z),z_n \right) \, ,
\end{split} \label{eq_residues}
\end{align}
where the first sum is over the roots of $B(z)$ whose imaginary part is strictly positive, and the second one is over the positive real roots of $B(z)$. We denote by ${\rm Res}(f(z), z_n)$ the residue of $f$ at point $z_n$. 

As an illustration, we consider the oscillatory incompressible case in the absence of relaxation ($\tilde\tau^{-1}\to0$), for which one has
\begin{align}
    R(z) = \frac{z}{z^2-\rmi \tilde\omega} \, ,
\end{align}
and thus for which $A(z)=z$ and $B(z)=z^2- \rmi \tilde\omega$ with the roots $z_{1,2}=\pm\sqrt{\rmi \tilde\omega}$. In this case, only the first term in the right-hand side of Eq.~\eqref{eq_residues} contributes, and since the Hankel function $H^{(1)}_0$ has no pole in $z_1=\sqrt{\rmi \tilde\omega}$, it yields
\begin{align}
    I\left[z/(z^2-\rmi \tilde\omega)\right] = \frac{\rmi \pi}{2} H^{(1)}_0(\sqrt{\rmi \tilde\omega}) = K_0(-\rmi \sqrt{\rmi \tilde\omega})  ~~  ,
\label{eq_analytic_incompressible}
\end{align}
where $K_\nu(x)$ is the $\nu^{\rm th}$ modified Bessel function of the second kind. An expansion of Eq.~\eqref{eq_analytic_incompressible} in series of $\tilde\omega$ yields the result given in Eq.~\eqref{eq_matrix_incompressible_noRelaxation}.

The same procedure can be applied for the compressible case, and was used to obtain Eqs.~\eqref{eq_analytical_steady} and \eqref{eq_analytic_frequency} of the main text.

\bibliography{biblio}

\begin{thebibliography}{77}%
\makeatletter
\providecommand \@ifxundefined [1]{%
 \@ifx{#1\undefined}
}%
\providecommand \@ifnum [1]{%
 \ifnum #1\expandafter \@firstoftwo
 \else \expandafter \@secondoftwo
 \fi
}%
\providecommand \@ifx [1]{%
 \ifx #1\expandafter \@firstoftwo
 \else \expandafter \@secondoftwo
 \fi
}%
\providecommand \natexlab [1]{#1}%
\providecommand \enquote  [1]{``#1''}%
\providecommand \bibnamefont  [1]{#1}%
\providecommand \bibfnamefont [1]{#1}%
\providecommand \citenamefont [1]{#1}%
\providecommand \href@noop [0]{\@secondoftwo}%
\providecommand \href [0]{\begingroup \@sanitize@url \@href}%
\providecommand \@href[1]{\@@startlink{#1}\@@href}%
\providecommand \@@href[1]{\endgroup#1\@@endlink}%
\providecommand \@sanitize@url [0]{\catcode `\\12\catcode `\$12\catcode
  `\&12\catcode `\#12\catcode `\^12\catcode `\_12\catcode `\%12\relax}%
\providecommand \@@startlink[1]{}%
\providecommand \@@endlink[0]{}%
\providecommand \url  [0]{\begingroup\@sanitize@url \@url }%
\providecommand \@url [1]{\endgroup\@href {#1}{\urlprefix }}%
\providecommand \urlprefix  [0]{URL }%
\providecommand \Eprint [0]{\href }%
\providecommand \doibase [0]{https://doi.org/}%
\providecommand \selectlanguage [0]{\@gobble}%
\providecommand \bibinfo  [0]{\@secondoftwo}%
\providecommand \bibfield  [0]{\@secondoftwo}%
\providecommand \translation [1]{[#1]}%
\providecommand \BibitemOpen [0]{}%
\providecommand \bibitemStop [0]{}%
\providecommand \bibitemNoStop [0]{.\EOS\space}%
\providecommand \EOS [0]{\spacefactor3000\relax}%
\providecommand \BibitemShut  [1]{\csname bibitem#1\endcsname}%
\let\auto@bib@innerbib\@empty
\bibitem [{\citenamefont {Pellegrino}\ \emph {et~al.}(2017)\citenamefont
  {Pellegrino}, \citenamefont {Torre},\ and\ \citenamefont
  {Polini}}]{pellegrino2017nonlocal}%
  \BibitemOpen
  \bibfield  {author} {\bibinfo {author} {\bibfnamefont {F.~M.~D.}\
  \bibnamefont {Pellegrino}}, \bibinfo {author} {\bibfnamefont
  {I.}~\bibnamefont {Torre}},\ and\ \bibinfo {author} {\bibfnamefont
  {M.}~\bibnamefont {Polini}},\ }\href
  {https://doi.org/10.1103/PhysRevB.96.195401} {\bibfield  {journal} {\bibinfo
  {journal} {Phys. Rev. B}\ }\textbf {\bibinfo {volume} {96}},\ \bibinfo
  {pages} {195401} (\bibinfo {year} {2017})}\BibitemShut {NoStop}%
\bibitem [{\citenamefont {Narozhny}\ and\ \citenamefont
  {Sch{\"u}tt}(2019)}]{narozhny2019magnetohydrodynamics}%
  \BibitemOpen
  \bibfield  {author} {\bibinfo {author} {\bibfnamefont {B.~N.}\ \bibnamefont
  {Narozhny}}\ and\ \bibinfo {author} {\bibfnamefont {M.}~\bibnamefont
  {Sch{\"u}tt}},\ }\href {https://doi.org/10.1103/PhysRevB.100.035125}
  {\bibfield  {journal} {\bibinfo  {journal} {Phys. Rev. B}\ }\textbf {\bibinfo
  {volume} {100}},\ \bibinfo {pages} {035125} (\bibinfo {year}
  {2019})}\BibitemShut {NoStop}%
\bibitem [{\citenamefont {Berdyugin}\ \emph {et~al.}(2019)\citenamefont
  {Berdyugin}, \citenamefont {Xu}, \citenamefont {Pellegrino}, \citenamefont
  {Krishna~Kumar}, \citenamefont {Principi}, \citenamefont {Torre},
  \citenamefont {Ben~Shalom}, \citenamefont {Taniguchi}, \citenamefont
  {Watanabe}, \citenamefont {Grigorieva}, \citenamefont {Polini}, \citenamefont
  {Geim},\ and\ \citenamefont {Bandurin}}]{berdyugin2019measuring}%
  \BibitemOpen
  \bibfield  {author} {\bibinfo {author} {\bibfnamefont {A.~I.}\ \bibnamefont
  {Berdyugin}}, \bibinfo {author} {\bibfnamefont {S.~G.}\ \bibnamefont {Xu}},
  \bibinfo {author} {\bibfnamefont {F.~M.~D.}\ \bibnamefont {Pellegrino}},
  \bibinfo {author} {\bibfnamefont {R.}~\bibnamefont {Krishna~Kumar}}, \bibinfo
  {author} {\bibfnamefont {A.}~\bibnamefont {Principi}}, \bibinfo {author}
  {\bibfnamefont {I.}~\bibnamefont {Torre}}, \bibinfo {author} {\bibfnamefont
  {M.}~\bibnamefont {Ben~Shalom}}, \bibinfo {author} {\bibfnamefont
  {T.}~\bibnamefont {Taniguchi}}, \bibinfo {author} {\bibfnamefont
  {K.}~\bibnamefont {Watanabe}}, \bibinfo {author} {\bibfnamefont {I.~V.}\
  \bibnamefont {Grigorieva}}, \bibinfo {author} {\bibfnamefont
  {M.}~\bibnamefont {Polini}}, \bibinfo {author} {\bibfnamefont {A.~K.}\
  \bibnamefont {Geim}},\ and\ \bibinfo {author} {\bibfnamefont {D.~A.}\
  \bibnamefont {Bandurin}},\ }\href {https://doi.org/10.1126/science.aau0685}
  {\bibfield  {journal} {\bibinfo  {journal} {Science}\ }\textbf {\bibinfo
  {volume} {364}},\ \bibinfo {pages} {162} (\bibinfo {year}
  {2019})}\BibitemShut {NoStop}%
\bibitem [{\citenamefont {Fossati}\ \emph {et~al.}(2022)\citenamefont
  {Fossati}, \citenamefont {Scheibner}, \citenamefont {Fruchart},\ and\
  \citenamefont {Vitelli}}]{fossati2022odd}%
  \BibitemOpen
  \bibfield  {author} {\bibinfo {author} {\bibfnamefont {M.}~\bibnamefont
  {Fossati}}, \bibinfo {author} {\bibfnamefont {C.}~\bibnamefont {Scheibner}},
  \bibinfo {author} {\bibfnamefont {M.}~\bibnamefont {Fruchart}},\ and\
  \bibinfo {author} {\bibfnamefont {V.}~\bibnamefont {Vitelli}},\ }\href@noop
  {} {\bibfield  {journal} {\bibinfo  {journal} {arXiv:2210.03669}\ } (\bibinfo
  {year} {2022})},\ \Eprint {https://arxiv.org/abs/2210.03669}
  {arXiv:2210.03669} \BibitemShut {NoStop}%
\bibitem [{\citenamefont {Green}\ \emph {et~al.}(2020)\citenamefont {Green},
  \citenamefont {Armas}, \citenamefont {{de Boer}},\ and\ \citenamefont
  {Giomi}}]{green2020topological}%
  \BibitemOpen
  \bibfield  {author} {\bibinfo {author} {\bibfnamefont {R.}~\bibnamefont
  {Green}}, \bibinfo {author} {\bibfnamefont {J.}~\bibnamefont {Armas}},
  \bibinfo {author} {\bibfnamefont {J.}~\bibnamefont {{de Boer}}},\ and\
  \bibinfo {author} {\bibfnamefont {L.}~\bibnamefont {Giomi}},\ }\href@noop {}
  {\bibfield  {journal} {\bibinfo  {journal} {arXiv:2011.12271}\ } (\bibinfo
  {year} {2020})},\ \Eprint {https://arxiv.org/abs/2011.12271}
  {arXiv:2011.12271} \BibitemShut {NoStop}%
\bibitem [{\citenamefont {Souslov}\ \emph {et~al.}(2019)\citenamefont
  {Souslov}, \citenamefont {Dasbiswas}, \citenamefont {Fruchart}, \citenamefont
  {Vaikuntanathan},\ and\ \citenamefont {Vitelli}}]{souslov2019topological}%
  \BibitemOpen
  \bibfield  {author} {\bibinfo {author} {\bibfnamefont {A.}~\bibnamefont
  {Souslov}}, \bibinfo {author} {\bibfnamefont {K.}~\bibnamefont {Dasbiswas}},
  \bibinfo {author} {\bibfnamefont {M.}~\bibnamefont {Fruchart}}, \bibinfo
  {author} {\bibfnamefont {S.}~\bibnamefont {Vaikuntanathan}},\ and\ \bibinfo
  {author} {\bibfnamefont {V.}~\bibnamefont {Vitelli}},\ }\href
  {https://doi.org/10.1103/PhysRevLett.122.128001} {\bibfield  {journal}
  {\bibinfo  {journal} {Phys. Rev. Lett.}\ }\textbf {\bibinfo {volume} {122}},\
  \bibinfo {pages} {128001} (\bibinfo {year} {2019})}\BibitemShut {NoStop}%
\bibitem [{\citenamefont {Tauber}\ \emph {et~al.}(2019)\citenamefont {Tauber},
  \citenamefont {Delplace},\ and\ \citenamefont
  {Venaille}}]{tauber2019bulkinterface}%
  \BibitemOpen
  \bibfield  {author} {\bibinfo {author} {\bibfnamefont {C.}~\bibnamefont
  {Tauber}}, \bibinfo {author} {\bibfnamefont {P.}~\bibnamefont {Delplace}},\
  and\ \bibinfo {author} {\bibfnamefont {A.}~\bibnamefont {Venaille}},\ }\href
  {https://doi.org/10.1017/jfm.2019.233} {\bibfield  {journal} {\bibinfo
  {journal} {J. Fluid Mech.}\ }\textbf {\bibinfo {volume} {868}},\ \bibinfo
  {pages} {R2} (\bibinfo {year} {2019})}\BibitemShut {NoStop}%
\bibitem [{\citenamefont {Banerjee}\ \emph {et~al.}(2017)\citenamefont
  {Banerjee}, \citenamefont {Souslov}, \citenamefont {Abanov},\ and\
  \citenamefont {Vitelli}}]{banerjee2017odd}%
  \BibitemOpen
  \bibfield  {author} {\bibinfo {author} {\bibfnamefont {D.}~\bibnamefont
  {Banerjee}}, \bibinfo {author} {\bibfnamefont {A.}~\bibnamefont {Souslov}},
  \bibinfo {author} {\bibfnamefont {A.~G.}\ \bibnamefont {Abanov}},\ and\
  \bibinfo {author} {\bibfnamefont {V.}~\bibnamefont {Vitelli}},\ }\href
  {https://doi.org/10.1038/s41467-017-01378-7} {\bibfield  {journal} {\bibinfo
  {journal} {Nat. Commun.}\ }\textbf {\bibinfo {volume} {8}},\ \bibinfo {pages}
  {1573} (\bibinfo {year} {2017})}\BibitemShut {NoStop}%
\bibitem [{\citenamefont {Abanov}\ \emph {et~al.}(2018)\citenamefont {Abanov},
  \citenamefont {Can},\ and\ \citenamefont {Ganeshan}}]{abanov2018odd}%
  \BibitemOpen
  \bibfield  {author} {\bibinfo {author} {\bibfnamefont {A.}~\bibnamefont
  {Abanov}}, \bibinfo {author} {\bibfnamefont {T.}~\bibnamefont {Can}},\ and\
  \bibinfo {author} {\bibfnamefont {S.}~\bibnamefont {Ganeshan}},\ }\href
  {https://doi.org/10.21468/SciPostPhys.5.1.010} {\bibfield  {journal}
  {\bibinfo  {journal} {SciPost Phys.}\ }\textbf {\bibinfo {volume} {5}},\
  \bibinfo {pages} {010} (\bibinfo {year} {2018})}\BibitemShut {NoStop}%
\bibitem [{\citenamefont {Ganeshan}\ and\ \citenamefont
  {Abanov}(2017)}]{ganeshan2017odd}%
  \BibitemOpen
  \bibfield  {author} {\bibinfo {author} {\bibfnamefont {S.}~\bibnamefont
  {Ganeshan}}\ and\ \bibinfo {author} {\bibfnamefont {A.~G.}\ \bibnamefont
  {Abanov}},\ }\href {https://doi.org/10.1103/PhysRevFluids.2.094101}
  {\bibfield  {journal} {\bibinfo  {journal} {Phys. Rev. Fluids}\ }\textbf
  {\bibinfo {volume} {2}},\ \bibinfo {pages} {094101} (\bibinfo {year}
  {2017})}\BibitemShut {NoStop}%
\bibitem [{\citenamefont {Khain}\ \emph {et~al.}(2022)\citenamefont {Khain},
  \citenamefont {Scheibner}, \citenamefont {Fruchart},\ and\ \citenamefont
  {Vitelli}}]{khain2022stokes}%
  \BibitemOpen
  \bibfield  {author} {\bibinfo {author} {\bibfnamefont {T.}~\bibnamefont
  {Khain}}, \bibinfo {author} {\bibfnamefont {C.}~\bibnamefont {Scheibner}},
  \bibinfo {author} {\bibfnamefont {M.}~\bibnamefont {Fruchart}},\ and\
  \bibinfo {author} {\bibfnamefont {V.}~\bibnamefont {Vitelli}},\ }\href
  {https://doi.org/10.1017/jfm.2021.1079} {\bibfield  {journal} {\bibinfo
  {journal} {J. Fluid Mech.}\ }\textbf {\bibinfo {volume} {934}},\ \bibinfo
  {pages} {A23} (\bibinfo {year} {2022})}\BibitemShut {NoStop}%
\bibitem [{\citenamefont {Scheibner}\ \emph {et~al.}(2020)\citenamefont
  {Scheibner}, \citenamefont {Souslov}, \citenamefont {Banerjee}, \citenamefont
  {Sur{\'o}wka}, \citenamefont {Irvine},\ and\ \citenamefont
  {Vitelli}}]{scheibner2020odd}%
  \BibitemOpen
  \bibfield  {author} {\bibinfo {author} {\bibfnamefont {C.}~\bibnamefont
  {Scheibner}}, \bibinfo {author} {\bibfnamefont {A.}~\bibnamefont {Souslov}},
  \bibinfo {author} {\bibfnamefont {D.}~\bibnamefont {Banerjee}}, \bibinfo
  {author} {\bibfnamefont {P.}~\bibnamefont {Sur{\'o}wka}}, \bibinfo {author}
  {\bibfnamefont {W.~T.~M.}\ \bibnamefont {Irvine}},\ and\ \bibinfo {author}
  {\bibfnamefont {V.}~\bibnamefont {Vitelli}},\ }\href
  {https://doi.org/10.1038/s41567-020-0795-y} {\bibfield  {journal} {\bibinfo
  {journal} {Nat. Phys.}\ }\textbf {\bibinfo {volume} {16}},\ \bibinfo {pages}
  {475} (\bibinfo {year} {2020})}\BibitemShut {NoStop}%
\bibitem [{\citenamefont {Braverman}\ \emph {et~al.}(2021)\citenamefont
  {Braverman}, \citenamefont {Scheibner}, \citenamefont {VanSaders},\ and\
  \citenamefont {Vitelli}}]{braverman2021topological}%
  \BibitemOpen
  \bibfield  {author} {\bibinfo {author} {\bibfnamefont {L.}~\bibnamefont
  {Braverman}}, \bibinfo {author} {\bibfnamefont {C.}~\bibnamefont
  {Scheibner}}, \bibinfo {author} {\bibfnamefont {B.}~\bibnamefont
  {VanSaders}},\ and\ \bibinfo {author} {\bibfnamefont {V.}~\bibnamefont
  {Vitelli}},\ }\href {https://doi.org/10.1103/PhysRevLett.127.268001}
  {\bibfield  {journal} {\bibinfo  {journal} {Phys. Rev. Lett.}\ }\textbf
  {\bibinfo {volume} {127}},\ \bibinfo {pages} {268001} (\bibinfo {year}
  {2021})}\BibitemShut {NoStop}%
\bibitem [{\citenamefont {Floyd}\ \emph {et~al.}(2022)\citenamefont {Floyd},
  \citenamefont {Dinner},\ and\ \citenamefont
  {Vaikuntanathan}}]{floyd2022signatures}%
  \BibitemOpen
  \bibfield  {author} {\bibinfo {author} {\bibfnamefont {C.}~\bibnamefont
  {Floyd}}, \bibinfo {author} {\bibfnamefont {A.~R.}\ \bibnamefont {Dinner}},\
  and\ \bibinfo {author} {\bibfnamefont {S.}~\bibnamefont {Vaikuntanathan}},\
  }\href@noop {} {\bibfield  {journal} {\bibinfo  {journal} {arXiv:2210.01159}\
  } (\bibinfo {year} {2022})},\ \Eprint {https://arxiv.org/abs/2210.01159}
  {arXiv:2210.01159} \BibitemShut {NoStop}%
\bibitem [{\citenamefont {Fruchart}\ \emph
  {et~al.}(2022{\natexlab{a}})\citenamefont {Fruchart}, \citenamefont
  {Scheibner},\ and\ \citenamefont {Vitelli}}]{fruchart2022odda}%
  \BibitemOpen
  \bibfield  {author} {\bibinfo {author} {\bibfnamefont {M.}~\bibnamefont
  {Fruchart}}, \bibinfo {author} {\bibfnamefont {C.}~\bibnamefont
  {Scheibner}},\ and\ \bibinfo {author} {\bibfnamefont {V.}~\bibnamefont
  {Vitelli}},\ }\href@noop {} {\bibfield  {journal} {\bibinfo  {journal}
  {arXiv:2207.00071}\ } (\bibinfo {year} {2022}{\natexlab{a}})},\ \Eprint
  {https://arxiv.org/abs/2207.00071} {arXiv:2207.00071} \BibitemShut {NoStop}%
\bibitem [{\citenamefont {Liao}\ \emph {et~al.}(2020)\citenamefont {Liao},
  \citenamefont {Irvine},\ and\ \citenamefont
  {Vaikuntanathan}}]{liao2020rectification}%
  \BibitemOpen
  \bibfield  {author} {\bibinfo {author} {\bibfnamefont {Z.}~\bibnamefont
  {Liao}}, \bibinfo {author} {\bibfnamefont {W.~T.~M.}\ \bibnamefont
  {Irvine}},\ and\ \bibinfo {author} {\bibfnamefont {S.}~\bibnamefont
  {Vaikuntanathan}},\ }\href {https://doi.org/10.1103/PhysRevX.10.021036}
  {\bibfield  {journal} {\bibinfo  {journal} {Phys. Rev. X}\ }\textbf {\bibinfo
  {volume} {10}},\ \bibinfo {pages} {021036} (\bibinfo {year}
  {2020})}\BibitemShut {NoStop}%
\bibitem [{\citenamefont {Epstein}\ and\ \citenamefont
  {Mandadapu}(2020)}]{epstein2020timereversal}%
  \BibitemOpen
  \bibfield  {author} {\bibinfo {author} {\bibfnamefont {J.~M.}\ \bibnamefont
  {Epstein}}\ and\ \bibinfo {author} {\bibfnamefont {K.~K.}\ \bibnamefont
  {Mandadapu}},\ }\href {https://doi.org/10.1103/PhysRevE.101.052614}
  {\bibfield  {journal} {\bibinfo  {journal} {Phys. Rev. E}\ }\textbf {\bibinfo
  {volume} {101}},\ \bibinfo {pages} {052614} (\bibinfo {year}
  {2020})}\BibitemShut {NoStop}%
\bibitem [{\citenamefont {Hargus}\ \emph {et~al.}(2020)\citenamefont {Hargus},
  \citenamefont {Klymko}, \citenamefont {Epstein},\ and\ \citenamefont
  {Mandadapu}}]{hargus2020time}%
  \BibitemOpen
  \bibfield  {author} {\bibinfo {author} {\bibfnamefont {C.}~\bibnamefont
  {Hargus}}, \bibinfo {author} {\bibfnamefont {K.}~\bibnamefont {Klymko}},
  \bibinfo {author} {\bibfnamefont {J.~M.}\ \bibnamefont {Epstein}},\ and\
  \bibinfo {author} {\bibfnamefont {K.~K.}\ \bibnamefont {Mandadapu}},\ }\href
  {https://doi.org/10.1063/5.0006441} {\bibfield  {journal} {\bibinfo
  {journal} {J. Chem. Phys.}\ }\textbf {\bibinfo {volume} {152}},\ \bibinfo
  {pages} {201102} (\bibinfo {year} {2020})}\BibitemShut {NoStop}%
\bibitem [{\citenamefont {Lier}\ \emph {et~al.}(2022)\citenamefont {Lier},
  \citenamefont {Armas}, \citenamefont {Bo}, \citenamefont {Duclut},
  \citenamefont {J{\"u}licher},\ and\ \citenamefont
  {Sur{\'o}wka}}]{lier2022passive}%
  \BibitemOpen
  \bibfield  {author} {\bibinfo {author} {\bibfnamefont {R.}~\bibnamefont
  {Lier}}, \bibinfo {author} {\bibfnamefont {J.}~\bibnamefont {Armas}},
  \bibinfo {author} {\bibfnamefont {S.}~\bibnamefont {Bo}}, \bibinfo {author}
  {\bibfnamefont {C.}~\bibnamefont {Duclut}}, \bibinfo {author} {\bibfnamefont
  {F.}~\bibnamefont {J{\"u}licher}},\ and\ \bibinfo {author} {\bibfnamefont
  {P.}~\bibnamefont {Sur{\'o}wka}},\ }\href
  {https://doi.org/10.1103/PhysRevE.105.054607} {\bibfield  {journal} {\bibinfo
   {journal} {Phys. Rev. E}\ }\textbf {\bibinfo {volume} {105}},\ \bibinfo
  {pages} {054607} (\bibinfo {year} {2022})}\BibitemShut {NoStop}%
\bibitem [{\citenamefont {Han}\ \emph {et~al.}(2021)\citenamefont {Han},
  \citenamefont {Fruchart}, \citenamefont {Scheibner}, \citenamefont
  {Vaikuntanathan}, \citenamefont {{de Pablo}},\ and\ \citenamefont
  {Vitelli}}]{han2021fluctuating}%
  \BibitemOpen
  \bibfield  {author} {\bibinfo {author} {\bibfnamefont {M.}~\bibnamefont
  {Han}}, \bibinfo {author} {\bibfnamefont {M.}~\bibnamefont {Fruchart}},
  \bibinfo {author} {\bibfnamefont {C.}~\bibnamefont {Scheibner}}, \bibinfo
  {author} {\bibfnamefont {S.}~\bibnamefont {Vaikuntanathan}}, \bibinfo
  {author} {\bibfnamefont {J.~J.}\ \bibnamefont {{de Pablo}}},\ and\ \bibinfo
  {author} {\bibfnamefont {V.}~\bibnamefont {Vitelli}},\ }\href
  {https://doi.org/10.1038/s41567-021-01360-7} {\bibfield  {journal} {\bibinfo
  {journal} {Nat. Phys.}\ }\textbf {\bibinfo {volume} {17}},\ \bibinfo {pages}
  {1260} (\bibinfo {year} {2021})}\BibitemShut {NoStop}%
\bibitem [{\citenamefont {Hargus}\ \emph {et~al.}(2021)\citenamefont {Hargus},
  \citenamefont {Epstein},\ and\ \citenamefont {Mandadapu}}]{hargus2021odd}%
  \BibitemOpen
  \bibfield  {author} {\bibinfo {author} {\bibfnamefont {C.}~\bibnamefont
  {Hargus}}, \bibinfo {author} {\bibfnamefont {J.~M.}\ \bibnamefont
  {Epstein}},\ and\ \bibinfo {author} {\bibfnamefont {K.~K.}\ \bibnamefont
  {Mandadapu}},\ }\href {https://doi.org/10.1103/PhysRevLett.127.178001}
  {\bibfield  {journal} {\bibinfo  {journal} {Phys. Rev. Lett.}\ }\textbf
  {\bibinfo {volume} {127}},\ \bibinfo {pages} {178001} (\bibinfo {year}
  {2021})}\BibitemShut {NoStop}%
\bibitem [{\citenamefont {Soni}\ \emph {et~al.}(2019)\citenamefont {Soni},
  \citenamefont {Bililign}, \citenamefont {Magkiriadou}, \citenamefont
  {Sacanna}, \citenamefont {Bartolo}, \citenamefont {Shelley},\ and\
  \citenamefont {Irvine}}]{soni2019odd}%
  \BibitemOpen
  \bibfield  {author} {\bibinfo {author} {\bibfnamefont {V.}~\bibnamefont
  {Soni}}, \bibinfo {author} {\bibfnamefont {E.~S.}\ \bibnamefont {Bililign}},
  \bibinfo {author} {\bibfnamefont {S.}~\bibnamefont {Magkiriadou}}, \bibinfo
  {author} {\bibfnamefont {S.}~\bibnamefont {Sacanna}}, \bibinfo {author}
  {\bibfnamefont {D.}~\bibnamefont {Bartolo}}, \bibinfo {author} {\bibfnamefont
  {M.~J.}\ \bibnamefont {Shelley}},\ and\ \bibinfo {author} {\bibfnamefont
  {W.~T.~M.}\ \bibnamefont {Irvine}},\ }\href
  {https://doi.org/10.1038/s41567-019-0603-8} {\bibfield  {journal} {\bibinfo
  {journal} {Nat. Phys.}\ }\textbf {\bibinfo {volume} {15}},\ \bibinfo {pages}
  {1188} (\bibinfo {year} {2019})}\BibitemShut {NoStop}%
\bibitem [{\citenamefont {Reichhardt}\ and\ \citenamefont
  {Reichhardt}(2019)}]{reichhardt2019active}%
  \BibitemOpen
  \bibfield  {author} {\bibinfo {author} {\bibfnamefont {C.}~\bibnamefont
  {Reichhardt}}\ and\ \bibinfo {author} {\bibfnamefont {C.~J.~O.}\ \bibnamefont
  {Reichhardt}},\ }\href {https://doi.org/10.1103/PhysRevE.100.012604}
  {\bibfield  {journal} {\bibinfo  {journal} {Phys. Rev. E}\ }\textbf {\bibinfo
  {volume} {100}},\ \bibinfo {pages} {012604} (\bibinfo {year}
  {2019})}\BibitemShut {NoStop}%
\bibitem [{\citenamefont {Markovich}\ and\ \citenamefont
  {Lubensky}(2021)}]{markovich2021odd}%
  \BibitemOpen
  \bibfield  {author} {\bibinfo {author} {\bibfnamefont {T.}~\bibnamefont
  {Markovich}}\ and\ \bibinfo {author} {\bibfnamefont {T.~C.}\ \bibnamefont
  {Lubensky}},\ }\href {https://doi.org/10.1103/PhysRevLett.127.048001}
  {\bibfield  {journal} {\bibinfo  {journal} {Phys. Rev. Lett.}\ }\textbf
  {\bibinfo {volume} {127}},\ \bibinfo {pages} {048001} (\bibinfo {year}
  {2021})}\BibitemShut {NoStop}%
\bibitem [{\citenamefont {Reichhardt}\ and\ \citenamefont
  {Reichhardt}(2022)}]{reichhardt2022active}%
  \BibitemOpen
  \bibfield  {author} {\bibinfo {author} {\bibfnamefont {C.~J.~O.}\
  \bibnamefont {Reichhardt}}\ and\ \bibinfo {author} {\bibfnamefont
  {C.}~\bibnamefont {Reichhardt}},\ }\href
  {https://doi.org/10.1209/0295-5075/ac2adc} {\bibfield  {journal} {\bibinfo
  {journal} {EPL}\ }\textbf {\bibinfo {volume} {137}},\ \bibinfo {pages}
  {66004} (\bibinfo {year} {2022})}\BibitemShut {NoStop}%
\bibitem [{\citenamefont {F{\"u}rthauer}\ \emph {et~al.}(2012)\citenamefont
  {F{\"u}rthauer}, \citenamefont {Strempel}, \citenamefont {Grill},\ and\
  \citenamefont {J{\"u}licher}}]{furthauer2012active}%
  \BibitemOpen
  \bibfield  {author} {\bibinfo {author} {\bibfnamefont {S.}~\bibnamefont
  {F{\"u}rthauer}}, \bibinfo {author} {\bibfnamefont {M.}~\bibnamefont
  {Strempel}}, \bibinfo {author} {\bibfnamefont {S.~W.}\ \bibnamefont
  {Grill}},\ and\ \bibinfo {author} {\bibfnamefont {F.}~\bibnamefont
  {J{\"u}licher}},\ }\href {https://doi.org/10.1140/epje/i2012-12089-6}
  {\bibfield  {journal} {\bibinfo  {journal} {Eur. Phys. J. E}\ }\textbf
  {\bibinfo {volume} {35}},\ \bibinfo {pages} {89} (\bibinfo {year}
  {2012})}\BibitemShut {NoStop}%
\bibitem [{\citenamefont {Tan}\ \emph {et~al.}(2022)\citenamefont {Tan},
  \citenamefont {Mietke}, \citenamefont {Li}, \citenamefont {Chen},
  \citenamefont {Higinbotham}, \citenamefont {Foster}, \citenamefont {Gokhale},
  \citenamefont {Dunkel},\ and\ \citenamefont {Fakhri}}]{tan2022odd}%
  \BibitemOpen
  \bibfield  {author} {\bibinfo {author} {\bibfnamefont {T.~H.}\ \bibnamefont
  {Tan}}, \bibinfo {author} {\bibfnamefont {A.}~\bibnamefont {Mietke}},
  \bibinfo {author} {\bibfnamefont {J.}~\bibnamefont {Li}}, \bibinfo {author}
  {\bibfnamefont {Y.}~\bibnamefont {Chen}}, \bibinfo {author} {\bibfnamefont
  {H.}~\bibnamefont {Higinbotham}}, \bibinfo {author} {\bibfnamefont {P.~J.}\
  \bibnamefont {Foster}}, \bibinfo {author} {\bibfnamefont {S.}~\bibnamefont
  {Gokhale}}, \bibinfo {author} {\bibfnamefont {J.}~\bibnamefont {Dunkel}},\
  and\ \bibinfo {author} {\bibfnamefont {N.}~\bibnamefont {Fakhri}},\ }\href
  {https://doi.org/10.1038/s41586-022-04889-6} {\bibfield  {journal} {\bibinfo
  {journal} {Nature}\ }\textbf {\bibinfo {volume} {607}},\ \bibinfo {pages}
  {287} (\bibinfo {year} {2022})}\BibitemShut {NoStop}%
\bibitem [{\citenamefont {Banerjee}\ \emph {et~al.}(2021)\citenamefont
  {Banerjee}, \citenamefont {Vitelli}, \citenamefont {J{\"u}licher},\ and\
  \citenamefont {Sur{\'o}wka}}]{banerjee2021active}%
  \BibitemOpen
  \bibfield  {author} {\bibinfo {author} {\bibfnamefont {D.}~\bibnamefont
  {Banerjee}}, \bibinfo {author} {\bibfnamefont {V.}~\bibnamefont {Vitelli}},
  \bibinfo {author} {\bibfnamefont {F.}~\bibnamefont {J{\"u}licher}},\ and\
  \bibinfo {author} {\bibfnamefont {P.}~\bibnamefont {Sur{\'o}wka}},\ }\href
  {https://doi.org/10.1103/PhysRevLett.126.138001} {\bibfield  {journal}
  {\bibinfo  {journal} {Phys. Rev. Lett.}\ }\textbf {\bibinfo {volume} {126}},\
  \bibinfo {pages} {138001} (\bibinfo {year} {2021})}\BibitemShut {NoStop}%
\bibitem [{\citenamefont {Avron}\ \emph {et~al.}(1995)\citenamefont {Avron},
  \citenamefont {Seiler},\ and\ \citenamefont {Zograf}}]{avron1995viscosity}%
  \BibitemOpen
  \bibfield  {author} {\bibinfo {author} {\bibfnamefont {J.~E.}\ \bibnamefont
  {Avron}}, \bibinfo {author} {\bibfnamefont {R.}~\bibnamefont {Seiler}},\ and\
  \bibinfo {author} {\bibfnamefont {P.~G.}\ \bibnamefont {Zograf}},\ }\href
  {https://doi.org/10.1103/PhysRevLett.75.697} {\bibfield  {journal} {\bibinfo
  {journal} {Phys. Rev. Lett.}\ }\textbf {\bibinfo {volume} {75}},\ \bibinfo
  {pages} {697} (\bibinfo {year} {1995})}\BibitemShut {NoStop}%
\bibitem [{\citenamefont {L{\'e}vay}(1995)}]{levay1995berry}%
  \BibitemOpen
  \bibfield  {author} {\bibinfo {author} {\bibfnamefont {P.}~\bibnamefont
  {L{\'e}vay}},\ }\href {https://doi.org/10.1063/1.531066} {\bibfield
  {journal} {\bibinfo  {journal} {J. Math. Phys.}\ }\textbf {\bibinfo {volume}
  {36}},\ \bibinfo {pages} {2792} (\bibinfo {year} {1995})}\BibitemShut
  {NoStop}%
\bibitem [{\citenamefont {Avron}(1998)}]{avron1998odd}%
  \BibitemOpen
  \bibfield  {author} {\bibinfo {author} {\bibfnamefont {J.~E.}\ \bibnamefont
  {Avron}},\ }\href {https://doi.org/10.1023/A:1023084404080} {\bibfield
  {journal} {\bibinfo  {journal} {J. Stat. Phys.}\ }\textbf {\bibinfo {volume}
  {92}},\ \bibinfo {pages} {543} (\bibinfo {year} {1998})}\BibitemShut
  {NoStop}%
\bibitem [{\citenamefont {Lucas}\ and\ \citenamefont
  {Sur{\'o}wka}(2014)}]{lucas2014phenomenology}%
  \BibitemOpen
  \bibfield  {author} {\bibinfo {author} {\bibfnamefont {A.}~\bibnamefont
  {Lucas}}\ and\ \bibinfo {author} {\bibfnamefont {P.}~\bibnamefont
  {Sur{\'o}wka}},\ }\href {https://doi.org/10.1103/PhysRevE.90.063005}
  {\bibfield  {journal} {\bibinfo  {journal} {Phys. Rev. E}\ }\textbf {\bibinfo
  {volume} {90}},\ \bibinfo {pages} {063005} (\bibinfo {year}
  {2014})}\BibitemShut {NoStop}%
\bibitem [{\citenamefont {Kogan}(2016)}]{kogan2016lift}%
  \BibitemOpen
  \bibfield  {author} {\bibinfo {author} {\bibfnamefont {E.}~\bibnamefont
  {Kogan}},\ }\href {https://doi.org/10.1103/PhysRevE.94.043111} {\bibfield
  {journal} {\bibinfo  {journal} {Phys. Rev. E}\ }\textbf {\bibinfo {volume}
  {94}},\ \bibinfo {pages} {043111} (\bibinfo {year} {2016})}\BibitemShut
  {NoStop}%
\bibitem [{\citenamefont {Alekseev}(2016)}]{alekseev2016negative}%
  \BibitemOpen
  \bibfield  {author} {\bibinfo {author} {\bibfnamefont {P.~S.}\ \bibnamefont
  {Alekseev}},\ }\href {https://doi.org/10.1103/PhysRevLett.117.166601}
  {\bibfield  {journal} {\bibinfo  {journal} {Phys. Rev. Lett.}\ }\textbf
  {\bibinfo {volume} {117}},\ \bibinfo {pages} {166601} (\bibinfo {year}
  {2016})}\BibitemShut {NoStop}%
\bibitem [{\citenamefont {Scaffidi}\ \emph {et~al.}(2017)\citenamefont
  {Scaffidi}, \citenamefont {Nandi}, \citenamefont {Schmidt}, \citenamefont
  {Mackenzie},\ and\ \citenamefont {Moore}}]{scaffidi2017hydrodynamic}%
  \BibitemOpen
  \bibfield  {author} {\bibinfo {author} {\bibfnamefont {T.}~\bibnamefont
  {Scaffidi}}, \bibinfo {author} {\bibfnamefont {N.}~\bibnamefont {Nandi}},
  \bibinfo {author} {\bibfnamefont {B.}~\bibnamefont {Schmidt}}, \bibinfo
  {author} {\bibfnamefont {A.~P.}\ \bibnamefont {Mackenzie}},\ and\ \bibinfo
  {author} {\bibfnamefont {J.~E.}\ \bibnamefont {Moore}},\ }\href
  {https://doi.org/10.1103/PhysRevLett.118.226601} {\bibfield  {journal}
  {\bibinfo  {journal} {Phys. Rev. Lett.}\ }\textbf {\bibinfo {volume} {118}},\
  \bibinfo {pages} {226601} (\bibinfo {year} {2017})}\BibitemShut {NoStop}%
\bibitem [{\citenamefont {Delacr{\'e}taz}\ and\ \citenamefont
  {Gromov}(2017)}]{delacretaz2017transport}%
  \BibitemOpen
  \bibfield  {author} {\bibinfo {author} {\bibfnamefont {L.~V.}\ \bibnamefont
  {Delacr{\'e}taz}}\ and\ \bibinfo {author} {\bibfnamefont {A.}~\bibnamefont
  {Gromov}},\ }\href {https://doi.org/10.1103/PhysRevLett.119.226602}
  {\bibfield  {journal} {\bibinfo  {journal} {Phys. Rev. Lett.}\ }\textbf
  {\bibinfo {volume} {119}},\ \bibinfo {pages} {226602} (\bibinfo {year}
  {2017})}\BibitemShut {NoStop}%
\bibitem [{\citenamefont {Gusev}\ \emph {et~al.}(2018)\citenamefont {Gusev},
  \citenamefont {Levin}, \citenamefont {Levinson},\ and\ \citenamefont
  {Bakarov}}]{gusev2018viscous}%
  \BibitemOpen
  \bibfield  {author} {\bibinfo {author} {\bibfnamefont {G.~M.}\ \bibnamefont
  {Gusev}}, \bibinfo {author} {\bibfnamefont {A.~D.}\ \bibnamefont {Levin}},
  \bibinfo {author} {\bibfnamefont {E.~V.}\ \bibnamefont {Levinson}},\ and\
  \bibinfo {author} {\bibfnamefont {A.~K.}\ \bibnamefont {Bakarov}},\ }\href
  {https://doi.org/10.1103/PhysRevB.98.161303} {\bibfield  {journal} {\bibinfo
  {journal} {Phys. Rev. B}\ }\textbf {\bibinfo {volume} {98}},\ \bibinfo
  {pages} {161303} (\bibinfo {year} {2018})}\BibitemShut {NoStop}%
\bibitem [{\citenamefont {Hosaka}\ \emph
  {et~al.}(2021{\natexlab{a}})\citenamefont {Hosaka}, \citenamefont {Komura},\
  and\ \citenamefont {Andelman}}]{hosaka2021nonreciprocal}%
  \BibitemOpen
  \bibfield  {author} {\bibinfo {author} {\bibfnamefont {Y.}~\bibnamefont
  {Hosaka}}, \bibinfo {author} {\bibfnamefont {S.}~\bibnamefont {Komura}},\
  and\ \bibinfo {author} {\bibfnamefont {D.}~\bibnamefont {Andelman}},\ }\href
  {https://doi.org/10.1103/PhysRevE.103.042610} {\bibfield  {journal} {\bibinfo
   {journal} {Phys. Rev. E}\ }\textbf {\bibinfo {volume} {103}},\ \bibinfo
  {pages} {042610} (\bibinfo {year} {2021}{\natexlab{a}})}\BibitemShut
  {NoStop}%
\bibitem [{\citenamefont {Hosaka}\ \emph
  {et~al.}(2021{\natexlab{b}})\citenamefont {Hosaka}, \citenamefont {Komura},\
  and\ \citenamefont {Andelman}}]{hosaka2021hydrodynamic}%
  \BibitemOpen
  \bibfield  {author} {\bibinfo {author} {\bibfnamefont {Y.}~\bibnamefont
  {Hosaka}}, \bibinfo {author} {\bibfnamefont {S.}~\bibnamefont {Komura}},\
  and\ \bibinfo {author} {\bibfnamefont {D.}~\bibnamefont {Andelman}},\ }\href
  {https://doi.org/10.1103/PhysRevE.104.064613} {\bibfield  {journal} {\bibinfo
   {journal} {Phys. Rev. E}\ }\textbf {\bibinfo {volume} {104}},\ \bibinfo
  {pages} {064613} (\bibinfo {year} {2021}{\natexlab{b}})}\BibitemShut
  {NoStop}%
\bibitem [{\citenamefont {Fruchart}\ \emph
  {et~al.}(2022{\natexlab{b}})\citenamefont {Fruchart}, \citenamefont {Han},
  \citenamefont {Scheibner},\ and\ \citenamefont {Vitelli}}]{fruchart2022odd}%
  \BibitemOpen
  \bibfield  {author} {\bibinfo {author} {\bibfnamefont {M.}~\bibnamefont
  {Fruchart}}, \bibinfo {author} {\bibfnamefont {M.}~\bibnamefont {Han}},
  \bibinfo {author} {\bibfnamefont {C.}~\bibnamefont {Scheibner}},\ and\
  \bibinfo {author} {\bibfnamefont {V.}~\bibnamefont {Vitelli}},\ }\href@noop
  {} {\bibfield  {journal} {\bibinfo  {journal} {arXiv:2202.02037}\ } (\bibinfo
  {year} {2022}{\natexlab{b}})},\ \Eprint {https://arxiv.org/abs/2202.02037}
  {arXiv:2202.02037} \BibitemShut {NoStop}%
\bibitem [{\citenamefont {Banerjee}\ \emph {et~al.}(2022)\citenamefont
  {Banerjee}, \citenamefont {Souslov},\ and\ \citenamefont
  {Vitelli}}]{banerjee2022hydrodynamic}%
  \BibitemOpen
  \bibfield  {author} {\bibinfo {author} {\bibfnamefont {D.}~\bibnamefont
  {Banerjee}}, \bibinfo {author} {\bibfnamefont {A.}~\bibnamefont {Souslov}},\
  and\ \bibinfo {author} {\bibfnamefont {V.}~\bibnamefont {Vitelli}},\ }\href
  {https://doi.org/10.1103/PhysRevFluids.7.043301} {\bibfield  {journal}
  {\bibinfo  {journal} {Phys. Rev. Fluids}\ }\textbf {\bibinfo {volume} {7}},\
  \bibinfo {pages} {043301} (\bibinfo {year} {2022})}\BibitemShut {NoStop}%
\bibitem [{\citenamefont {Bureau}\ \emph {et~al.}(2022)\citenamefont {Bureau},
  \citenamefont {Coupier},\ and\ \citenamefont {Salez}}]{bureau2022lift}%
  \BibitemOpen
  \bibfield  {author} {\bibinfo {author} {\bibfnamefont {L.}~\bibnamefont
  {Bureau}}, \bibinfo {author} {\bibfnamefont {G.}~\bibnamefont {Coupier}},\
  and\ \bibinfo {author} {\bibfnamefont {T.}~\bibnamefont {Salez}},\ }\bibfield
   {journal} {\bibinfo  {journal} {arXiv:2207.04538}\ }\href
  {https://doi.org/10.48550/ARXIV.2207.04538} {10.48550/ARXIV.2207.04538}
  (\bibinfo {year} {2022}),\ \Eprint {https://arxiv.org/abs/2207.04538}
  {arXiv:2207.04538} \BibitemShut {NoStop}%
\bibitem [{\citenamefont {Lamb}(1932)}]{lamb1932hydrodynamics}%
  \BibitemOpen
  \bibfield  {author} {\bibinfo {author} {\bibfnamefont {H.}~\bibnamefont
  {Lamb}},\ }\href@noop {} {\emph {\bibinfo {title} {Hydrodynamics}}}\
  (\bibinfo  {publisher} {{Cambridge University Press}},\ \bibinfo {address}
  {{New York, U.S.A.}},\ \bibinfo {year} {1932})\BibitemShut {NoStop}%
\bibitem [{\citenamefont {Oseen}(1910)}]{oseen1910uber}%
  \BibitemOpen
  \bibfield  {author} {\bibinfo {author} {\bibfnamefont {C.~W.}\ \bibnamefont
  {Oseen}},\ }\href@noop {} {\bibfield  {journal} {\bibinfo  {journal} {Ark.
  Mat., Astron. Fys.}\ }\textbf {\bibinfo {volume} {6}},\ \bibinfo {pages} {1}
  (\bibinfo {year} {1910})}\BibitemShut {NoStop}%
\bibitem [{\citenamefont {Proudman}\ and\ \citenamefont
  {Pearson}(1957)}]{proudman1957expansions}%
  \BibitemOpen
  \bibfield  {author} {\bibinfo {author} {\bibfnamefont {I.}~\bibnamefont
  {Proudman}}\ and\ \bibinfo {author} {\bibfnamefont {J.~R.~A.}\ \bibnamefont
  {Pearson}},\ }\href {https://doi.org/10.1017/S0022112057000105} {\bibfield
  {journal} {\bibinfo  {journal} {J. Fluid Mech.}\ }\textbf {\bibinfo {volume}
  {2}},\ \bibinfo {pages} {237} (\bibinfo {year} {1957})}\BibitemShut {NoStop}%
\bibitem [{\citenamefont {Chester}\ \emph {et~al.}(1969)\citenamefont
  {Chester}, \citenamefont {Breach},\ and\ \citenamefont
  {Proudman}}]{chester_breach_proudman_1969}%
  \BibitemOpen
  \bibfield  {author} {\bibinfo {author} {\bibfnamefont {W.}~\bibnamefont
  {Chester}}, \bibinfo {author} {\bibfnamefont {D.~R.}\ \bibnamefont
  {Breach}},\ and\ \bibinfo {author} {\bibfnamefont {I.}~\bibnamefont
  {Proudman}},\ }\href {https://doi.org/10.1017/S0022112069000851} {\bibfield
  {journal} {\bibinfo  {journal} {Journal of Fluid Mechanics}\ }\textbf
  {\bibinfo {volume} {37}},\ \bibinfo {pages} {751–760} (\bibinfo {year}
  {1969})}\BibitemShut {NoStop}%
\bibitem [{\citenamefont {Veysey}\ and\ \citenamefont
  {Goldenfeld}(2007)}]{veysey2007simple}%
  \BibitemOpen
  \bibfield  {author} {\bibinfo {author} {\bibfnamefont {J.}~\bibnamefont
  {Veysey}}\ and\ \bibinfo {author} {\bibfnamefont {N.}~\bibnamefont
  {Goldenfeld}},\ }\href {https://doi.org/10.1103/RevModPhys.79.883} {\bibfield
   {journal} {\bibinfo  {journal} {Rev. Mod. Phys.}\ }\textbf {\bibinfo
  {volume} {79}},\ \bibinfo {pages} {883} (\bibinfo {year} {2007})}\BibitemShut
  {NoStop}%
\bibitem [{\citenamefont {Rosenhead}(1963)}]{rosenhead1963laminar}%
  \BibitemOpen
  \bibfield  {author} {\bibinfo {author} {\bibfnamefont {L.}~\bibnamefont
  {Rosenhead}},\ }\href@noop {} {\emph {\bibinfo {title} {Laminar Boundary
  Layers}}}\ (\bibinfo  {publisher} {{Clarendon Press}},\ \bibinfo {address}
  {{Oxford, England}},\ \bibinfo {year} {1963})\BibitemShut {NoStop}%
\bibitem [{\citenamefont {Happel}\ and\ \citenamefont
  {Brenner}(1983)}]{happel1983low}%
  \BibitemOpen
  \bibfield  {author} {\bibinfo {author} {\bibfnamefont {J.}~\bibnamefont
  {Happel}}\ and\ \bibinfo {author} {\bibfnamefont {H.}~\bibnamefont
  {Brenner}},\ }\href@noop {} {\emph {\bibinfo {title} {Low {{Reynolds}} Number
  Hydrodynamics}}}\ (\bibinfo  {publisher} {{Springer}},\ \bibinfo {address}
  {{The Hague, The Netherlands}},\ \bibinfo {year} {1983})\BibitemShut
  {NoStop}%
\bibitem [{\citenamefont {Van~Dyke}(1975)}]{vandyke1975perturbation}%
  \BibitemOpen
  \bibfield  {author} {\bibinfo {author} {\bibfnamefont {M.}~\bibnamefont
  {Van~Dyke}},\ }\href@noop {} {\emph {\bibinfo {title} {Perturbation Methods
  in Fluid Mechanics}}}\ (\bibinfo  {publisher} {{Parabolic Press}},\ \bibinfo
  {address} {{Stanford, California}},\ \bibinfo {year} {1975})\BibitemShut
  {NoStop}%
\bibitem [{\citenamefont {Saffman}\ and\ \citenamefont
  {Delbr{\"u}ck}(1975)}]{saffman1975brownian}%
  \BibitemOpen
  \bibfield  {author} {\bibinfo {author} {\bibfnamefont {P.~G.}\ \bibnamefont
  {Saffman}}\ and\ \bibinfo {author} {\bibfnamefont {M.}~\bibnamefont
  {Delbr{\"u}ck}},\ }\href {https://doi.org/10.1073/pnas.72.8.3111} {\bibfield
  {journal} {\bibinfo  {journal} {Proc. Natl. Acad. Sci. U.S.A.}\ }\textbf
  {\bibinfo {volume} {72}},\ \bibinfo {pages} {3111} (\bibinfo {year}
  {1975})}\BibitemShut {NoStop}%
\bibitem [{\citenamefont {Saffman}(1976)}]{saffman1976brownian}%
  \BibitemOpen
  \bibfield  {author} {\bibinfo {author} {\bibfnamefont {P.~G.}\ \bibnamefont
  {Saffman}},\ }\href {https://doi.org/10.1017/S0022112076001511} {\bibfield
  {journal} {\bibinfo  {journal} {J. Fluid Mech.}\ }\textbf {\bibinfo {volume}
  {73}},\ \bibinfo {pages} {593} (\bibinfo {year} {1976})}\BibitemShut
  {NoStop}%
\bibitem [{\citenamefont {Seki}\ and\ \citenamefont
  {Komura}(1993)}]{seki1993brownian}%
  \BibitemOpen
  \bibfield  {author} {\bibinfo {author} {\bibfnamefont {K.}~\bibnamefont
  {Seki}}\ and\ \bibinfo {author} {\bibfnamefont {S.}~\bibnamefont {Komura}},\
  }\href {https://doi.org/10.1103/PhysRevE.47.2377} {\bibfield  {journal}
  {\bibinfo  {journal} {Phys. Rev. E}\ }\textbf {\bibinfo {volume} {47}},\
  \bibinfo {pages} {2377} (\bibinfo {year} {1993})}\BibitemShut {NoStop}%
\bibitem [{\citenamefont {Levine}\ and\ \citenamefont
  {Lubensky}(2001)}]{levine2001response}%
  \BibitemOpen
  \bibfield  {author} {\bibinfo {author} {\bibfnamefont {A.~J.}\ \bibnamefont
  {Levine}}\ and\ \bibinfo {author} {\bibfnamefont {T.~C.}\ \bibnamefont
  {Lubensky}},\ }\href {https://doi.org/10.1103/PhysRevE.63.041510} {\bibfield
  {journal} {\bibinfo  {journal} {Phys. Rev. E}\ }\textbf {\bibinfo {volume}
  {63}},\ \bibinfo {pages} {041510} (\bibinfo {year} {2001})}\BibitemShut
  {NoStop}%
\bibitem [{\citenamefont {Camley}\ and\ \citenamefont
  {Brown}(2011)}]{camley2011creeping}%
  \BibitemOpen
  \bibfield  {author} {\bibinfo {author} {\bibfnamefont {B.~A.}\ \bibnamefont
  {Camley}}\ and\ \bibinfo {author} {\bibfnamefont {F.~L.~H.}\ \bibnamefont
  {Brown}},\ }\href {https://doi.org/10.1103/PhysRevE.84.021904} {\bibfield
  {journal} {\bibinfo  {journal} {Phys. Rev. E}\ }\textbf {\bibinfo {volume}
  {84}},\ \bibinfo {pages} {021904} (\bibinfo {year} {2011})}\BibitemShut
  {NoStop}%
\bibitem [{\citenamefont {Levine}\ and\ \citenamefont
  {MacKintosh}(2002)}]{levine2002dynamics}%
  \BibitemOpen
  \bibfield  {author} {\bibinfo {author} {\bibfnamefont {A.~J.}\ \bibnamefont
  {Levine}}\ and\ \bibinfo {author} {\bibfnamefont {F.~C.}\ \bibnamefont
  {MacKintosh}},\ }\href {https://doi.org/10.1103/PhysRevE.66.061606}
  {\bibfield  {journal} {\bibinfo  {journal} {Phys. Rev. E}\ }\textbf {\bibinfo
  {volume} {66}},\ \bibinfo {pages} {061606} (\bibinfo {year}
  {2002})}\BibitemShut {NoStop}%
\bibitem [{\citenamefont {Toner}(2012)}]{toner2012birth}%
  \BibitemOpen
  \bibfield  {author} {\bibinfo {author} {\bibfnamefont {J.}~\bibnamefont
  {Toner}},\ }\href {https://doi.org/10.1103/PhysRevLett.108.088102} {\bibfield
   {journal} {\bibinfo  {journal} {Phys. Rev. Lett.}\ }\textbf {\bibinfo
  {volume} {108}},\ \bibinfo {pages} {088102} (\bibinfo {year}
  {2012})}\BibitemShut {NoStop}%
\bibitem [{\citenamefont {Chen}\ \emph {et~al.}(2020)\citenamefont {Chen},
  \citenamefont {Lee},\ and\ \citenamefont {Toner}}]{chen2020moving}%
  \BibitemOpen
  \bibfield  {author} {\bibinfo {author} {\bibfnamefont {L.}~\bibnamefont
  {Chen}}, \bibinfo {author} {\bibfnamefont {C.~F.}\ \bibnamefont {Lee}},\ and\
  \bibinfo {author} {\bibfnamefont {J.}~\bibnamefont {Toner}},\ }\href
  {https://doi.org/10.1103/PhysRevLett.125.098003} {\bibfield  {journal}
  {\bibinfo  {journal} {Phys. Rev. Lett.}\ }\textbf {\bibinfo {volume} {125}},\
  \bibinfo {pages} {098003} (\bibinfo {year} {2020})}\BibitemShut {NoStop}%
\bibitem [{\citenamefont {Ranft}\ \emph {et~al.}(2010)\citenamefont {Ranft},
  \citenamefont {Basan}, \citenamefont {Elgeti}, \citenamefont {Joanny},
  \citenamefont {Prost},\ and\ \citenamefont
  {J{\"u}licher}}]{ranft2010fluidization}%
  \BibitemOpen
  \bibfield  {author} {\bibinfo {author} {\bibfnamefont {J.}~\bibnamefont
  {Ranft}}, \bibinfo {author} {\bibfnamefont {M.}~\bibnamefont {Basan}},
  \bibinfo {author} {\bibfnamefont {J.}~\bibnamefont {Elgeti}}, \bibinfo
  {author} {\bibfnamefont {J.-F.}\ \bibnamefont {Joanny}}, \bibinfo {author}
  {\bibfnamefont {J.}~\bibnamefont {Prost}},\ and\ \bibinfo {author}
  {\bibfnamefont {F.}~\bibnamefont {J{\"u}licher}},\ }\href
  {https://doi.org/10.1073/pnas.1011086107} {\bibfield  {journal} {\bibinfo
  {journal} {Proc. Natl. Acad. Sci. U.S.A.}\ }\textbf {\bibinfo {volume}
  {107}},\ \bibinfo {pages} {20863} (\bibinfo {year} {2010})}\BibitemShut
  {NoStop}%
\bibitem [{\citenamefont {Ben Ali~Zinati}\ \emph {et~al.}(2022)\citenamefont
  {Ben Ali~Zinati}, \citenamefont {Duclut}, \citenamefont {Mahdisoltani},
  \citenamefont {Gambassi},\ and\ \citenamefont
  {Golestanian}}]{benalizinati2022stochastic}%
  \BibitemOpen
  \bibfield  {author} {\bibinfo {author} {\bibfnamefont {R.}~\bibnamefont {Ben
  Ali~Zinati}}, \bibinfo {author} {\bibfnamefont {C.}~\bibnamefont {Duclut}},
  \bibinfo {author} {\bibfnamefont {S.}~\bibnamefont {Mahdisoltani}}, \bibinfo
  {author} {\bibfnamefont {A.}~\bibnamefont {Gambassi}},\ and\ \bibinfo
  {author} {\bibfnamefont {R.}~\bibnamefont {Golestanian}},\ }\href
  {https://doi.org/10.1209/0295-5075/ac48c9} {\bibfield  {journal} {\bibinfo
  {journal} {EPL}\ }\textbf {\bibinfo {volume} {136}},\ \bibinfo {pages}
  {50003} (\bibinfo {year} {2022})}\BibitemShut {NoStop}%
\bibitem [{\citenamefont {Barentin}\ \emph {et~al.}(2000)\citenamefont
  {Barentin}, \citenamefont {Muller}, \citenamefont {Ybert}, \citenamefont
  {Joanny},\ and\ \citenamefont {{di Meglio}}}]{barentin2000shear}%
  \BibitemOpen
  \bibfield  {author} {\bibinfo {author} {\bibfnamefont {C.}~\bibnamefont
  {Barentin}}, \bibinfo {author} {\bibfnamefont {P.}~\bibnamefont {Muller}},
  \bibinfo {author} {\bibfnamefont {C.}~\bibnamefont {Ybert}}, \bibinfo
  {author} {\bibfnamefont {J.-F.}\ \bibnamefont {Joanny}},\ and\ \bibinfo
  {author} {\bibfnamefont {J.-M.}\ \bibnamefont {{di Meglio}}},\ }\href
  {https://doi.org/10.1007/s101890050049} {\bibfield  {journal} {\bibinfo
  {journal} {Eur. Phys. J. E}\ }\textbf {\bibinfo {volume} {2}},\ \bibinfo
  {pages} {153} (\bibinfo {year} {2000})}\BibitemShut {NoStop}%
\bibitem [{\citenamefont {Elfring}\ \emph {et~al.}(2016)\citenamefont
  {Elfring}, \citenamefont {Leal},\ and\ \citenamefont
  {Squires}}]{elfring2016surface}%
  \BibitemOpen
  \bibfield  {author} {\bibinfo {author} {\bibfnamefont {G.~J.}\ \bibnamefont
  {Elfring}}, \bibinfo {author} {\bibfnamefont {L.~G.}\ \bibnamefont {Leal}},\
  and\ \bibinfo {author} {\bibfnamefont {T.~M.}\ \bibnamefont {Squires}},\
  }\href {https://doi.org/10.1017/jfm.2016.96} {\bibfield  {journal} {\bibinfo
  {journal} {J. Fluid Mech.}\ }\textbf {\bibinfo {volume} {792}},\ \bibinfo
  {pages} {712} (\bibinfo {year} {2016})}\BibitemShut {NoStop}%
\bibitem [{\citenamefont {Petroff}\ \emph {et~al.}(2015)\citenamefont
  {Petroff}, \citenamefont {Wu},\ and\ \citenamefont
  {Libchaber}}]{petroff2015fastmoving}%
  \BibitemOpen
  \bibfield  {author} {\bibinfo {author} {\bibfnamefont {A.~P.}\ \bibnamefont
  {Petroff}}, \bibinfo {author} {\bibfnamefont {X.-L.}\ \bibnamefont {Wu}},\
  and\ \bibinfo {author} {\bibfnamefont {A.}~\bibnamefont {Libchaber}},\ }\href
  {https://doi.org/10.1103/PhysRevLett.114.158102} {\bibfield  {journal}
  {\bibinfo  {journal} {Phys. Rev. Lett.}\ }\textbf {\bibinfo {volume} {114}},\
  \bibinfo {pages} {158102} (\bibinfo {year} {2015})}\BibitemShut {NoStop}%
\bibitem [{\citenamefont {Riedel}\ \emph {et~al.}(2005)\citenamefont {Riedel},
  \citenamefont {Kruse},\ and\ \citenamefont
  {Howard}}]{riedel2005selforganized}%
  \BibitemOpen
  \bibfield  {author} {\bibinfo {author} {\bibfnamefont {I.~H.}\ \bibnamefont
  {Riedel}}, \bibinfo {author} {\bibfnamefont {K.}~\bibnamefont {Kruse}},\ and\
  \bibinfo {author} {\bibfnamefont {J.}~\bibnamefont {Howard}},\ }\href
  {https://doi.org/10.1126/science.1110329} {\bibfield  {journal} {\bibinfo
  {journal} {Science}\ }\textbf {\bibinfo {volume} {309}},\ \bibinfo {pages}
  {300} (\bibinfo {year} {2005})}\BibitemShut {NoStop}%
\bibitem [{\citenamefont {J{\"u}licher}\ \emph {et~al.}(2018)\citenamefont
  {J{\"u}licher}, \citenamefont {Grill},\ and\ \citenamefont
  {Salbreux}}]{julicher2018hydrodynamic}%
  \BibitemOpen
  \bibfield  {author} {\bibinfo {author} {\bibfnamefont {F.}~\bibnamefont
  {J{\"u}licher}}, \bibinfo {author} {\bibfnamefont {S.~W.}\ \bibnamefont
  {Grill}},\ and\ \bibinfo {author} {\bibfnamefont {G.}~\bibnamefont
  {Salbreux}},\ }\href {https://doi.org/10.1088/1361-6633/aab6bb} {\bibfield
  {journal} {\bibinfo  {journal} {Rep. Prog. Phys.}\ }\textbf {\bibinfo
  {volume} {81}},\ \bibinfo {pages} {076601} (\bibinfo {year}
  {2018})}\BibitemShut {NoStop}%
\bibitem [{\citenamefont {MacKintosh}\ and\ \citenamefont
  {Lubensky}(1991)}]{mackintosh1991orientational}%
  \BibitemOpen
  \bibfield  {author} {\bibinfo {author} {\bibfnamefont {F.~C.}\ \bibnamefont
  {MacKintosh}}\ and\ \bibinfo {author} {\bibfnamefont {T.~C.}\ \bibnamefont
  {Lubensky}},\ }\href {https://doi.org/10.1103/PhysRevLett.67.1169} {\bibfield
   {journal} {\bibinfo  {journal} {Phys. Rev. Lett.}\ }\textbf {\bibinfo
  {volume} {67}},\ \bibinfo {pages} {1169} (\bibinfo {year}
  {1991})}\BibitemShut {NoStop}%
\bibitem [{\citenamefont {Williams}\ and\ \citenamefont
  {Hussey}(1972)}]{williams1972oscillating}%
  \BibitemOpen
  \bibfield  {author} {\bibinfo {author} {\bibfnamefont {R.~E.}\ \bibnamefont
  {Williams}}\ and\ \bibinfo {author} {\bibfnamefont {R.~G.}\ \bibnamefont
  {Hussey}},\ }\href {https://doi.org/10.1063/1.1693839} {\bibfield  {journal}
  {\bibinfo  {journal} {Phys. Fluids}\ }\textbf {\bibinfo {volume} {15}},\
  \bibinfo {pages} {2083} (\bibinfo {year} {1972})}\BibitemShut {NoStop}%
\bibitem [{\citenamefont {Dolfo}\ \emph {et~al.}(2020)\citenamefont {Dolfo},
  \citenamefont {Vigu{\'e}},\ and\ \citenamefont
  {Lhuillier}}]{dolfo2020stokes}%
  \BibitemOpen
  \bibfield  {author} {\bibinfo {author} {\bibfnamefont {G.}~\bibnamefont
  {Dolfo}}, \bibinfo {author} {\bibfnamefont {J.}~\bibnamefont {Vigu{\'e}}},\
  and\ \bibinfo {author} {\bibfnamefont {D.}~\bibnamefont {Lhuillier}},\
  }\href@noop {} {\bibfield  {journal} {\bibinfo  {journal} {arXiv:2011.12000}\
  } (\bibinfo {year} {2020})},\ \Eprint {https://arxiv.org/abs/2011.12000}
  {arXiv:2011.12000} \BibitemShut {NoStop}%
\bibitem [{\citenamefont {Lin}(2013{\natexlab{a}})}]{lin2013infinitea}%
  \BibitemOpen
  \bibfield  {author} {\bibinfo {author} {\bibfnamefont {Q.-G.}\ \bibnamefont
  {Lin}},\ }\href {https://doi.org/10.1080/10652469.2012.758119} {\bibfield
  {journal} {\bibinfo  {journal} {Integral Transforms and Special Functions}\
  }\textbf {\bibinfo {volume} {24}},\ \bibinfo {pages} {783} (\bibinfo {year}
  {2013}{\natexlab{a}})}\BibitemShut {NoStop}%
\bibitem [{\citenamefont {Masoud}\ and\ \citenamefont
  {Stone}(2019)}]{masoud2019reciprocal}%
  \BibitemOpen
  \bibfield  {author} {\bibinfo {author} {\bibfnamefont {H.}~\bibnamefont
  {Masoud}}\ and\ \bibinfo {author} {\bibfnamefont {H.~A.}\ \bibnamefont
  {Stone}},\ }\href {https://doi.org/10.1017/jfm.2019.553} {\bibfield
  {journal} {\bibinfo  {journal} {J. Fluid Mech.}\ }\textbf {\bibinfo {volume}
  {879}},\ \bibinfo {pages} {P1} (\bibinfo {year} {2019})}\BibitemShut
  {NoStop}%
\bibitem [{\citenamefont {Hosaka}\ \emph {et~al.}(2023)\citenamefont {Hosaka},
  \citenamefont {Golestanian},\ and\ \citenamefont
  {Vilfan}}]{hosaka2023lorentz}%
  \BibitemOpen
  \bibfield  {author} {\bibinfo {author} {\bibfnamefont {Y.}~\bibnamefont
  {Hosaka}}, \bibinfo {author} {\bibfnamefont {R.}~\bibnamefont
  {Golestanian}},\ and\ \bibinfo {author} {\bibfnamefont {A.}~\bibnamefont
  {Vilfan}},\ }\href@noop {} {\bibinfo {title} {Lorentz reciprocal theorem in
  fluids with odd viscosity}} (\bibinfo {year} {2023}),\ \Eprint
  {https://arxiv.org/abs/2305.15379} {arXiv:2305.15379 [cond-mat.soft]}
  \BibitemShut {NoStop}%
\bibitem [{\citenamefont {Weisenborn}\ and\ \citenamefont
  {Mazur}(1984)}]{WEISENBORN1984191}%
  \BibitemOpen
  \bibfield  {author} {\bibinfo {author} {\bibfnamefont {A.}~\bibnamefont
  {Weisenborn}}\ and\ \bibinfo {author} {\bibfnamefont {P.}~\bibnamefont
  {Mazur}},\ }\href
  {https://doi.org/https://doi.org/10.1016/0378-4371(84)90111-0} {\bibfield
  {journal} {\bibinfo  {journal} {Physica A: Statistical Mechanics and its
  Applications}\ }\textbf {\bibinfo {volume} {123}},\ \bibinfo {pages} {191}
  (\bibinfo {year} {1984})}\BibitemShut {NoStop}%
\bibitem [{\citenamefont {Korneev}\ \emph {et~al.}(2021)\citenamefont
  {Korneev}, \citenamefont {Kharzeev},\ and\ \citenamefont
  {Abanov}}]{korneev2021chiral}%
  \BibitemOpen
  \bibfield  {author} {\bibinfo {author} {\bibfnamefont {L.~A.}\ \bibnamefont
  {Korneev}}, \bibinfo {author} {\bibfnamefont {D.~E.}\ \bibnamefont
  {Kharzeev}},\ and\ \bibinfo {author} {\bibfnamefont {A.~G.}\ \bibnamefont
  {Abanov}},\ }\href {https://doi.org/10.1063/5.0058581} {\bibfield  {journal}
  {\bibinfo  {journal} {Phys. Fluids}\ }\textbf {\bibinfo {volume} {33}},\
  \bibinfo {pages} {083110} (\bibinfo {year} {2021})}\BibitemShut {NoStop}%
\bibitem [{\citenamefont {Evans}\ and\ \citenamefont
  {Sackmann}(1988)}]{evans1988translational}%
  \BibitemOpen
  \bibfield  {author} {\bibinfo {author} {\bibfnamefont {E.}~\bibnamefont
  {Evans}}\ and\ \bibinfo {author} {\bibfnamefont {E.}~\bibnamefont
  {Sackmann}},\ }\href {https://doi.org/10.1017/S0022112088003106} {\bibfield
  {journal} {\bibinfo  {journal} {J. Fluid Mech.}\ }\textbf {\bibinfo {volume}
  {194}},\ \bibinfo {pages} {553} (\bibinfo {year} {1988})}\BibitemShut
  {NoStop}%
\bibitem [{\citenamefont {Ramachandran}\ \emph {et~al.}(2010)\citenamefont
  {Ramachandran}, \citenamefont {Komura}, \citenamefont {Imai},\ and\
  \citenamefont {Seki}}]{ramachandran2010drag}%
  \BibitemOpen
  \bibfield  {author} {\bibinfo {author} {\bibfnamefont {S.}~\bibnamefont
  {Ramachandran}}, \bibinfo {author} {\bibfnamefont {S.}~\bibnamefont
  {Komura}}, \bibinfo {author} {\bibfnamefont {M.}~\bibnamefont {Imai}},\ and\
  \bibinfo {author} {\bibfnamefont {K.}~\bibnamefont {Seki}},\ }\href
  {https://doi.org/10.1140/epje/i2010-10577-3} {\bibfield  {journal} {\bibinfo
  {journal} {Eur. Phys. J. E}\ }\textbf {\bibinfo {volume} {31}},\ \bibinfo
  {pages} {303} (\bibinfo {year} {2010})}\BibitemShut {NoStop}%
\bibitem [{\citenamefont {Lucas}(2017)}]{lucas2017stokes}%
  \BibitemOpen
  \bibfield  {author} {\bibinfo {author} {\bibfnamefont {A.}~\bibnamefont
  {Lucas}},\ }\href {https://doi.org/10.1103/PhysRevB.95.115425} {\bibfield
  {journal} {\bibinfo  {journal} {Phys. Rev. B}\ }\textbf {\bibinfo {volume}
  {95}},\ \bibinfo {pages} {115425} (\bibinfo {year} {2017})}\BibitemShut
  {NoStop}%
\bibitem [{\citenamefont {Lin}(2013{\natexlab{b}})}]{lin2013infinite}%
  \BibitemOpen
  \bibfield  {author} {\bibinfo {author} {\bibfnamefont {Q.-G.}\ \bibnamefont
  {Lin}},\ }\href {https://doi.org/10.1080/10652469.2012.758119} {\bibfield
  {journal} {\bibinfo  {journal} {Integral Transforms Spec. Funct.}\ }\textbf
  {\bibinfo {volume} {24}},\ \bibinfo {pages} {783} (\bibinfo {year}
  {2013}{\natexlab{b}})}\BibitemShut {NoStop}%
\end{thebibliography}%

\end{document}